\PassOptionsToPackage{pdfpagelabels=false}{hyperref} 

\documentclass[fleqn,usenatbib]{mnras}

\usepackage{newtxtext,newtxmath}

\usepackage[T1]{fontenc}

\DeclareRobustCommand{\VAN}[3]{#2}
\let\VANthebibliography\thebibliography
\def\thebibliography{\DeclareRobustCommand{\VAN}[3]{##3}\VANthebibliography}

\usepackage{lineno}

\usepackage{graphicx}	
\usepackage{amsmath}	
\usepackage{svg}
\usepackage[flushleft]{threeparttable}

\title[SNe Ia within DES galaxy clusters]{Rates and properties of type Ia supernovae in galaxy clusters within the Dark Energy Survey}

\author[DES Collaboration]{
\parbox{\textwidth}{
\Large
M.~Toy,$^{1}$\thanks{E-mail: M.Toy@soton.ac.uk}
P.~Wiseman,$^{1}$
M.~Sullivan,$^{1}$
C.~Frohmaier,$^{1}$
O.~Graur,$^{2,3}$
A.~Palmese,$^{4}$
B.~Popovic,$^{5}$
T.~M.~Davis,$^{6}$
L.~Galbany,$^{7,8}$
L.~Kelsey,$^{2}$
C.~Lidman,$^{9,10}$
D.~Scolnic,$^{5}$
S.~Allam,$^{11}$
S.~Desai,$^{12}$
T.~M.~C.~Abbott,$^{13}$
M.~Aguena,$^{14}$
O.~Alves,$^{15}$
J.~Annis,$^{11}$
D.~Bacon,$^{2}$
E.~Bertin,$^{16,17}$
D.~Brooks,$^{18}$
D.~L.~Burke,$^{19,20}$
A.~Carnero~Rosell,$^{21,14,22}$
M.~Carrasco~Kind,$^{23,24}$
J.~Carretero,$^{25}$
F.~J.~Castander,$^{7,8}$
C.~Conselice,$^{26,27}$
L.~N.~da Costa,$^{14}$
M.~E.~S.~Pereira,$^{28}$
J.~De~Vicente,$^{29}$
H.~T.~Diehl,$^{11}$
P.~Doel,$^{18}$
S.~Everett,$^{30}$
I.~Ferrero,$^{31}$
J.~Frieman,$^{11,32}$
D.~W.~Gerdes,$^{33,15}$
D.~Gruen,$^{34}$
R.~A.~Gruendl,$^{23,24}$
G.~Gutierrez,$^{11}$
S.~R.~Hinton,$^{6}$
D.~L.~Hollowood,$^{35}$
K.~Honscheid,$^{36,37}$
D.~J.~James,$^{38}$
K.~Kuehn,$^{39,40}$
N.~Kuropatkin,$^{11}$
J.~L.~Marshall,$^{41}$
P.~Melchior,$^{42}$
J. Mena-Fern{\'a}ndez,$^{29}$
F.~Menanteau,$^{23,24}$
R.~Miquel,$^{43,25}$
A.~Pieres,$^{14,44}$
A.~A.~Plazas~Malag\'on,$^{19,20}$
A.~K.~Romer,$^{45}$
E.~Sanchez,$^{29}$
V.~Scarpine,$^{11}$
I.~Sevilla-Noarbe,$^{29}$
M.~Smith,$^{46}$
M.~Soares-Santos,$^{15}$
E.~Suchyta,$^{47}$
G.~Tarle,$^{15}$
C.~To,$^{36}$
and N.~Weaverdyck$^{15,48}$
\begin{center} (DES Collaboration) \end{center}
\begin{center} (Affiliations may be found at the end of the paper) \end{center}
}
\vspace{0.4cm}
\\
}

\date{Accepted XXX. Received YYY; in original form ZZZ}

\pubyear{2023}
\usepackage{eso-pic}

\AddToShipoutPictureBG*{
  \AtPageUpperLeft{
    \hspace{0.75\paperwidth}
    \raisebox{-5.5\baselineskip}{
      \makebox[0pt][l]{\textnormal{DES-2022-0695}}
}}}

\AddToShipoutPictureBG*{
  \AtPageUpperLeft{
    \hspace{0.75\paperwidth}
    \raisebox{-6.5\baselineskip}{
      \makebox[0pt][l]{\textnormal{FERMILAB-PUB-23-053-PPD}}
}}}

\begin{document}
\label{firstpage}
\pagerange{\pageref{firstpage}--\pageref{lastpage}}
\maketitle

\begin{abstract}
We identify 66 photometrically classified type Ia supernovae (SNe Ia) from the Dark Energy Survey (DES) that have occurred within red-sequence selected galaxy clusters. We compare light-curve and host galaxy properties of the cluster SNe to 1024 DES SNe Ia located in field galaxies, the largest comparison of two such samples at high redshift (z > 0.1). We find that cluster SN light curves decline faster than those in the field (97.7 per cent confidence). However, when limiting these samples to host galaxies of similar colour and mass, there is no significant difference in the SN light curve properties. Motivated by previous detections of a higher-normalised SN Ia delay time distribution in galaxy clusters, we measure the intrinsic rate of SNe Ia in cluster and field environments. We find the average ratio of the SN Ia rate per galaxy between high mass ($10\leq\log\mathrm{(M_{*}/M_{\odot})} \leq 11.25$) cluster and field galaxies to be $0.594 \pm0.068$. This difference is mass-dependent, with the ratio declining with increasing mass, which suggests that the stellar populations in cluster hosts are older than those in field hosts. We show that the mass-normalised rate (or SNe per unit mass) in massive-passive galaxies is consistent between cluster and field environments. Additionally, both of these rates are consistent with rates previously measured in clusters at similar redshifts. We conclude that in massive-passive galaxies, which are the dominant hosts of cluster SNe, the cluster DTD is comparable to the field.

\end{abstract}


\section{Introduction}
\label{sec:intro}
Type Ia supernovae (SNe Ia) are important transients in cosmology, as they can be used as standardisable candles to accurately measure distances, and have provided direct evidence for the accelerating expansion of our Universe \citep{Riess1998, Perl1999}. They are believed to be the explosions of carbon-oxygen white dwarf stars in binary systems, where an interaction with the companion star triggers a runaway thermonuclear reaction. 
The photometric properties of the resulting SNe show a dependence on the stellar populations of the galaxies where the SN occurred, and the properties of these galaxies in turn depend upon their large-scale environment: galaxies located within galaxy clusters are often older, and with less ongoing star formation than similar mass galaxies located in the field \citep{Bower1990}. In this paper, we investigate the properties of SNe Ia within galaxy clusters, and compare them to properties of SNe Ia located in field galaxies, to understand how the higher density environment affects SNe Ia.

SNe Ia are 
standardised as distance estimators using photomteric properties such as light-curve width (\lq stretch\rq, or $x_1$) and optical colour ($c$). Empirical relationships between peak brightness and light-curve width \citep{Rust1974,Pskovskii1977,Phillips1993} and colour \citep{Tripp1998} allow distances to SNe Ia to be estimated with a 6--7 per cent accuracy. These photometric SN Ia properties vary according to host galaxy properties of the SN, with those such as age \citep{2000AJ....120.1479H}, dust \citep{RiessDust,brout2020its}, stellar mass \citep {Sullivan_2010, Kelly2010} and specific star formation rate \citep[sSFR;][]{Sullivan_2010, Rigault2020} influencing the SN light curve properties. For example, galaxies with low star-formation rates host faster-declining (lower $x_1$) and fainter SNe Ia than similar mass galaxies with more vigorous star-formation rates \citep{Sullivan2006, Lampeitl2010}. 

Galaxy clusters are the largest gravitationally bound structures in our universe. Most of the galaxies and stars within clusters formed at least a few Gyrs ago \citep{Guglielmo_2015}. Compared to similar mass galaxies outside of clusters (field galaxies), cluster galaxies often have little ongoing star formation  \citep{ClusterSFR2, Haines2015}. Clusters contain up to a thousand galaxies within radii of up to 2.5\,Mpc, and as such these galaxies are more densely packed than those in the field. Due to this, environmental quenching of star formation in galaxies often occurs within clusters and groups. There is also evidence that the quenching of star formation in galaxies within clusters depends more strongly on the radial distance from the cluster centre (with cluster centres being the most efficiently quenched) than on the galaxy stellar mass \citep{van_der_Burg_2018}, leading to a greater number of lower mass galaxies with extinguished star formation \citep{van_der_Burg_2013} than the field.

The fraction of redder galaxy types increases with galaxy density in clusters \citep{Dressler1980}. Along with its effect on light-curve properties, galaxy colour is important for SN Ia cosmology, since there is evidence that SNe Ia in redder galaxies have a larger root mean square (r.m.s) \lq scatter\rq\ in their residuals on the distance--redshift ``Hubble diagram" \citep{Rigault2013, Kelsey2021}. Recently, \citet{Conor_2023} constructed a low-redshift sample of SNe Ia in galaxy clusters, and found evidence that SNe Ia in cluster galaxies differ from those in the field, even when accounting for the differing cluster and field galaxy populations. It is thus important to quantify the effects of cluster environment on SN Ia properties, both for the understanding of SN--host galaxy correlations as well as for precision cosmology.
 
Further to light curve differences and Hubble diagram scatter is an apparent offset in the SN Ia delay-time distribution (DTD) between clusters and field galaxies. The DTD describes the probability distribution of the time delay between the formation of a SN Ia progenitor star and its explosion as a SN, and is normalised 2--3 times higher in clusters compared to the field \citep{Moaz2017, Freundlich2021}. Constraining the cluster DTD requires measuring the SN Ia rate in clusters at a range of redshifts. A summary of many such measurements that have been inter-calibrated can be found in \citet{Freundlich2021}. 

In this paper, we undertake the largest study of high-redshift ($z > 0.1$) SNe Ia in galaxy clusters to date. We compare their light curve and host galaxy properties to a large sample of SNe Ia hosted in field galaxies. We also estimate the rates of both cluster and field SNe Ia in similar mass galaxies to determine if differences in environment leads to different rates. 

The paper is laid out as follows: Section \ref{sec:Method} describes the datasets used in the project, and outlines the selection cuts that are apply to our SN samples. It also describes the method for separating the samples into cluster SNe Ia and field SNe Ia. Section \ref{sec:Initial Results after Data Split} presents the light-curve properties of both field and cluster SNe Ia. The rates of SNe Ia per stellar mass for our samples are calculated in Section \ref{sec:Rates}. Section \ref{sec:Conclusions} summarises the paper. We assume a flat $\Lambda\mathrm{CDM}$ cosmology with a Hubble constant of $\mathrm{H_{0}}=70~\mathrm{km~s}^{-1}~\mathrm{Mpc} ^{-1}$, and $\Omega_\mathrm{M} = 0.3$.

\section{Dataset and Methodology}
\label{sec:Method}
We begin by describing the SN and galaxy cluster catalogues that we use, and present the method for identifying the SNe that occurred within galaxy clusters.

\subsection{The DES dataset}
\label{sec:Dataset}

The Dark Energy Survey (DES) was a survey that imaged 5000\,deg$^{2}$ of the southern sky in the $grizY$ bands. In this paper we make use of the Dark Energy Survey SN Programme \citep[DES-SN;][]{Bernstein2012} 5-year, photometrically-classified SN Ia sample \citep{Moller2022}. Clusters were identified within the first annual reduction of the science verification data (SVA1; \citealt{Rykoff_2016}) using the red-sequence Matched-filter Probabilistic Percolation (redMaPPer, henceforth RM) cluster finding algorithm \citep{Rykoff_2014}. The SV data encompasses $250$\,deg$^{2}$, including the SN fields, and was collected between November 2012 and February 2013. These images were reduced by an early version of the DES Data Management Pipeline \citep[DESDM:][]{2011arXiv1109.6741S, 2012SPIE.8451E..0DM, Desai2012}, covered the DES-SN fields at the depth of the full wide-field survey. It is a well-tested data set \citep{Bonnet2016,Jarvis2016,Jeffery2018} \footnote{\url{http://des.ncsa.illinois.edu/releases/sva1}}.

\subsubsection{Supernova Data}
\label{SNe Data}

DES-SN ran for five years for five-month seasons each year, using the Dark Energy Camera \citep[DECam:][]{Flaugher2015} to observe 27 deg$^2$ split over ten fields in the southern sky. These fields were repeatedly observed in the $griz$ filters, with an average of seven days between observations. Eight of the fields are \lq shallow\rq\ (single-visit depth of $m\sim23.5$\,mag) and two are \lq deep\rq\ (single-visit depth of $m\sim24.5$\,mag). The images were processed by the final DESDM pipeline \citep{Morganson_2018}, and transients identified using a difference imaging pipeline \citep{Kessler_2015}. Imaging artefacts were rejected using a machine-learning algorithm \citep{Goldstein_2015}, leaving around 30,000 candidate transients. These transients were then matched to a host galaxy using the directional light radius \citep[DLR;][]{Sullivan2006} method and deep galaxy images from \citet{Wiseman2020}.

We use the SN sample described in \citet{Moller2022}, where the photometric SN classifier SuperNNova \citep[SNN;][]{Moller_2019} was run on the DES 5-year candidate SN sample. We remove objects from our sample that have a SN Ia probability of $<50$ per cent. Candidates classified by SNN as SNe Ia are then fit with the SALT3 spectral energy distribution (SED) model \citep{SALT3} in the SuperNova Analysis framework \citep[\textsc{snana};][]{SNANA}. We apply a similar light curve selection to those described in \citet{vincenzi2021}, and used in \citet{Wiseman_2021}, on SN light-curve width (SALT3 $x_1$) and SN rest-frame colour (SALT3 $c$): $-3\leq x_{1} \leq3$ and $-0.3\leq c \leq 0.3$. These selection cuts reduce contamination from core-collapse SNe \citep{vincenzi2021}. We also require that each SN's host galaxy has a measured spectroscopic redshift \citep[see][]{vincenzi2021}, many of which were measured by the OzDES survey \citep{OzDES1, OzDES2}. 

The SN host galaxy properties are estimated following the method of \citet{Sullivan2006} and \citet{Smith_2020}, which used the P\'{E}GASE.2 \citep{Pegase1,Pegase2} spectral synthesis code. P\'{E}GASE.2 is used to generate synthetic galaxy spectral energy distributions (SEDs), which are then fitted to the host galaxy photometry obtained from the deep galaxy images \citep{Wiseman2020}. This results in a best-fitting stellar mass ($M_*$) and star-formation rate (SFR). We finally adjust the best-fitting template SED to exactly match the observed galaxy photometry to allow an accurate estimation of the galaxy rest-frame $UBVR$ magnitudes \citep[e.g.,][]{Kelsey2021}. For host NIR data, we make use of VISTA and DES imaging that have been combined for a subset of the DES-SN survey area \citep{Hartley2022}. 
\subsubsection{Galaxy Clusters}
\label{Cluster Data}

redMaPPer (RM) is a photometric red-sequence cluster finder, designed specifically for large scale surveys such as DES \citep{Rykoff_2014}. This red-sequence technique is built around richness estimators that have been optimised in \citet{2009ApJ...703..601R,2012ApJ...746..178R}. RM handles broad ranges of redshift well, and is ideal for use on DES data \citep{Rykoff_2016}.

For each cluster candidate, RM provides a richness estimate $\lambda$, and a scaling factor $S$ that accounts for survey incompleteness. These parameters are calculated such that each cluster with richness $\lambda$ has  $\lambda/S$ galaxies brighter than the limiting magnitude of the survey within the geometric survey mask.

We select all clusters with $\lambda/S \geq5$ from the SVA1 Gold 1.0.2 catalogue. While this catalogue is less reliable for analysis than other RM catalogues with more stringent richness cuts, it gives us a higher space density of clusters. Using the more stringent catalogues could cause us to misclassify SNe that occurred within less rich clusters as field SNe. For example, restricting our cluster catalogue to $\lambda \geq 20S$ only identifies 15 SNe within clusters. As such we do not use this richness cut, so we should be less likely to classify cluster SNe as field SNe. We investigate the effects of using a more conservative richness cut on our results in Appendix \ref{Richness_cut_appendix}, where we find that our results are broadly unchanged when using a richness cut of $\lambda \geq 15$. The $\lambda \geq 5S$ catalogue contains roughly $1000$ clusters within the DES-SN fields and spans a redshift range of $0.1 \leq z \leq 0.95$.

\subsection{Finding Supernovae Within Clusters}
\label{sec:Determining Supernovae Location}

To identify if a SN event occurred within a cluster, we follow the procedure outlined in \citet{Xavier}. Firstly, we check if any given SN was projected onto a cluster in the RM catalogue. For a given SN $s$ to be projected onto a cluster $k$ it must obey
\begin{equation}
    \label{projection}
    \cos\delta_s \cos\delta_k \cos(\alpha_s - \alpha_k) + \sin\delta_s \sin\delta_k \geq \cos\left(\theta_{\mathrm{max}}^{(k)}\right),
\end{equation}
where 
\begin{equation}
    \theta^{(k)}_{\mathrm{max}} \equiv \frac{1.5~\mathrm{Mpc}~(1+z_k)}{c\int_0^{z_k}\frac{\mathrm{d}z}{H(z)}}
\end{equation}
is the angular radius of cluster $k$, which we limit to a maximum value of 1.5\,Mpc, $c$ is the speed of light, $\alpha_{s(k)}$ and $\delta_{s(k)}$ are the right ascensions and declinations of the SN and cluster respectively, $z_k$ is the cluster redshift and $H(z)$ is the Hubble parameter. We use a 1.5\,Mpc limit to be consistent with other cluster SN rates in the literature \citep{Mannucci2008}, with a significant over-density of galaxies still present at these radii \citep{Hansen_2005}.

This matching identifies SNe that are projected onto the cluster, so we next compare their redshifts to determine if they are co-located. As galaxies within clusters are gravitationally bound, any measured redshift differences between cluster member galaxies arise from peculiar velocities and measurement uncertainties in the redshifts themselves.

We find the probability, $p$, for the redshift difference between a projected SN (with spectroscopic redshift) and cluster (with a photometric redshift) to be consistent with the SN being within the cluster. We assume the SNe and clusters have redshifts that are described by Gaussian probability distributions centred on the measured redshift, with standard deviations (uncertainties) $\sigma_s$ and $\sigma_k$ respectively. The probability for compatible redshifts is then
\begin{equation}
    p = \frac{1}{\sqrt{2\pi\left(\sigma^2_s+\sigma^2_k\right)}} \int_{-z_d}^{z_d} \mathrm{e}^{-\frac{[z-(z_s-z_k)]^2}{2\left(\sigma_s^2 + \sigma_k^2\right)}}.
\end{equation}

For our samples, the typical value of $\sigma_s$ is $\simeq0.001$, and the performance of the photometric redshifts for the clusters using RM is $\sigma_k/(1+z_k)\sim0.01$ \citep{Rykoff_2016}. As our cluster sample uses photometric redshifts, we set the maximum redshift difference, $z_d$, at 0.03 following \citet{Xavier}, who found this value to maximise the statistical difference between the cluster and field samples for photometric redshifts. Additionally, they calculate their contamination of field SNe in the cluster sample for their photometric sample at 42 per cent, with a combined photometric and spectroscopic contamination of 29 per cent. Using a similar calculation, we find our contamination to be $\sim$28 per cent, comparable to the combined contamination found in the \citet{Xavier} analysis, but smaller than their photometric samples contamination by 14 per cent. 

\begin{table}
\centering
\caption{The purity of a given cluster in a richness bin, estimated using \citet{Hao2010}.}
\begin{tabular}{|c|c|}
\hline
Richness & \begin{tabular}[c]{@{}c@{}}Purity\end{tabular} \\ \hline
5-10     & 0.60                                                            \\ 
10-15    & 0.75                                                           \\ 
15+      & 1.00                                                            \\ 
\hline
\end{tabular}
\label{tab:Purities}
\end{table}

The purity of our cluster sample is the probability that any given cluster is classified correctly, i.e., the probability that any identified cluster is real. We refer to these purities as $q$. While measurements of purity are not available for the SVA1 RM catalogue, we use estimates based on cluster richness from a similar cluster catalogue \citep{Hao2010}, and apply these to our sample. These purity estimates vary with richness, and are shown in Table \ref{tab:Purities}. We then make our final selection on the data. SNe that are projected onto a given cluster, with a combined probability of matching that cluster's redshift and the cluster itself being correctly identified of above 50 per cent ($pq > 0.5$), are considered as cluster SNe.

SNe Ia that are located within $1.5\,\mathrm{Mpc} \leq r \leq 2.5\,\mathrm{Mpc}$ of a given cluster have an uncertain cluster membership, and thus could contaminate our field sample with possible cluster SNe, and vice versa. We remove these SNe from both our samples if they pass the $pq$ test outlined above.

\begin{figure}
    \centering
	\includegraphics[width=\columnwidth]{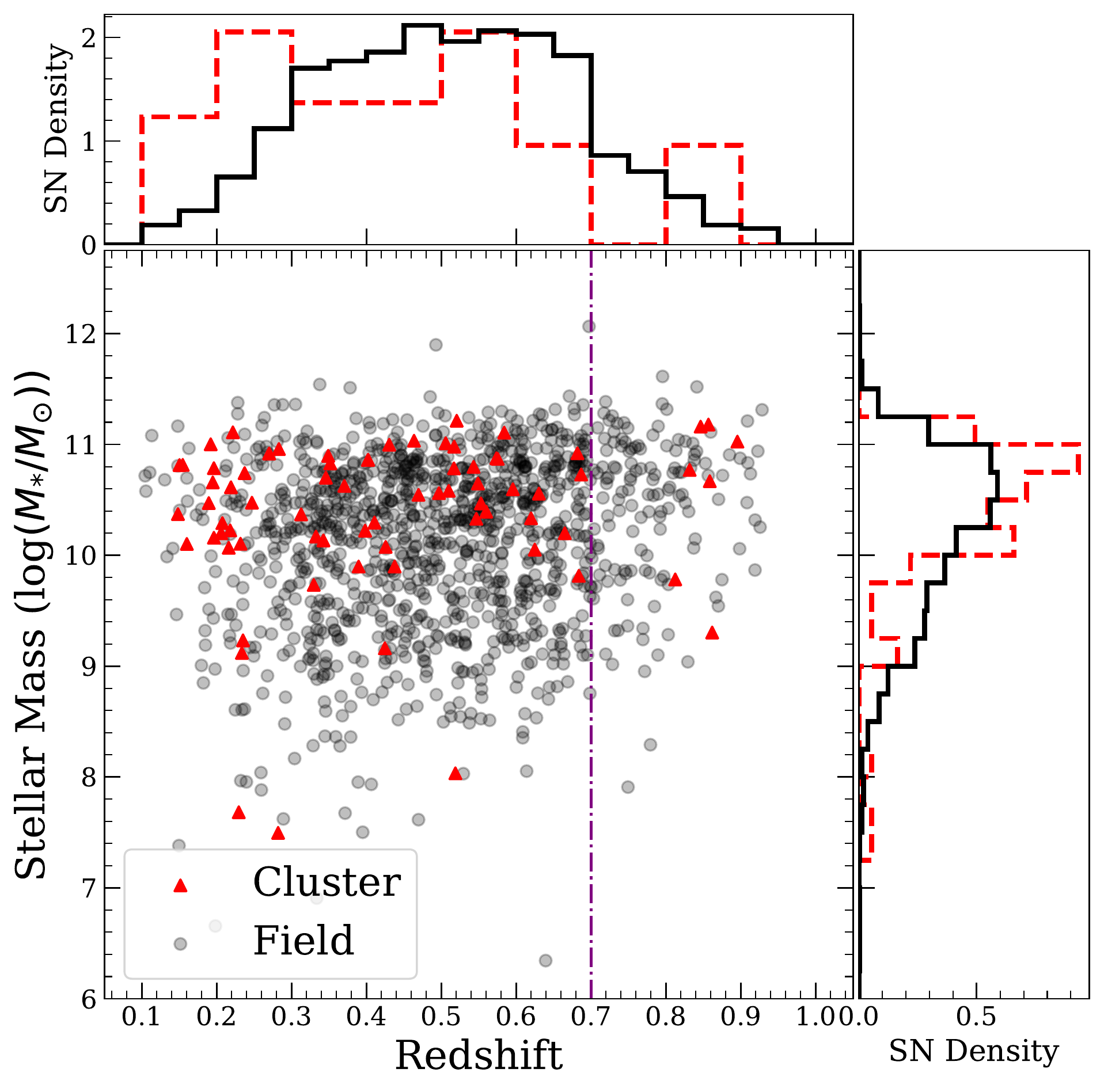}
	\caption{SN Ia host galaxy stellar mass versus redshift for our two SN Ia samples (cluster SNe as red triangles, and field SNe as grey circles). Average uncertainties on the host masses are <$\sim$ 0.04\,dex. We note a lack of SNe Ia in field and cluster galaxies at $z>0.7$ and $\log\left(M/M_{\odot}\right) < 10$. As such we make a redshift selection of $z < 0.7$, shown by the dashed line.}
    \label{fig:zcut}
\end{figure}

Fig.~\ref{fig:zcut} shows the redshift distribution of our two samples, together with the stellar masses of the identified SN Ia host galaxies. We find a lack of SN Ia hosts in both field and clusters around and below $\log(M/M_{\odot}) = 10$ at z $\geq$ 0.7. This lack of lower mass host galaxies is likely a selection effect (the lower mass galaxies are fainter and therefore harder to measure a spectroscopic redshift for) and we make a selection in redshift of $z<0.7$ for both samples. This leaves 66 SNe Ia located within clusters, and 1024 SNe Ia located in the field. Table~\ref{cuts} shows how many SNe are removed at each stage of our selection. 

\begin{figure}
    \centering
	\includegraphics[width=\columnwidth]{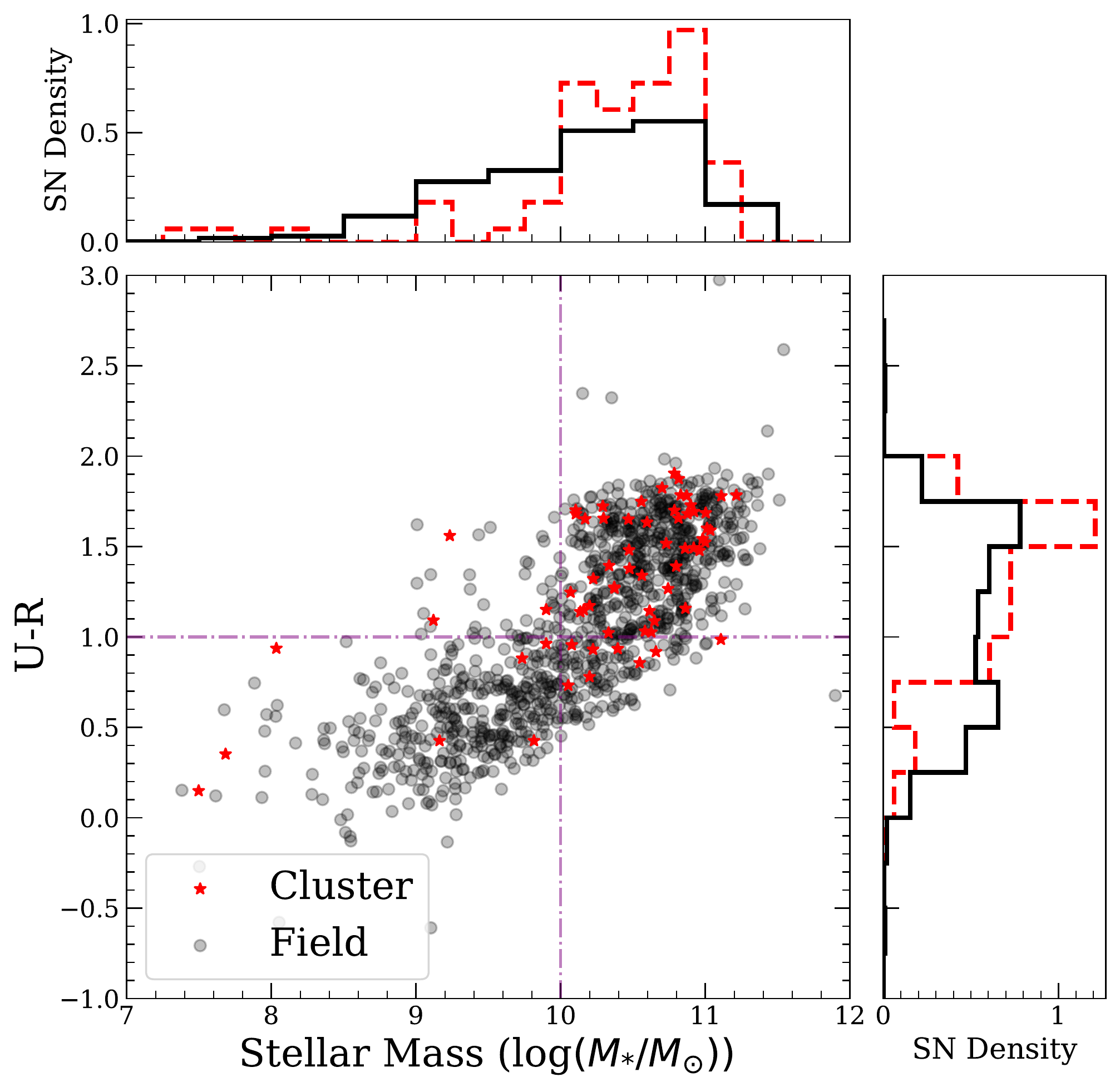}
    \caption{$U-R$ rest-frame colour versus stellar mass for SN Ia host galaxies. Cluster host galaxies are redder and more massive on average than those in the field. The horizontal and vertical  lines indicate our selection of red, massive galaxies, with masses $\log\left(M/M_{\odot}\right) > 10$ and $U-R > 1$.}
    \label{fig:U_R_Host}
\end{figure}

We show our $U-R$ vs stellar mass distribution in Fig.~\ref{fig:U_R_Host}. It is evident that cluster and field SNe inhabit different galaxy populations. To fairly compare the samples like-for-like, we select a sub-sample of cluster and field SNe Ia with similar host properties, and investigate any host dependencies. We select SNe with a host galaxy rest-frame $U-R$ colour of $> 1$ and a host stellar mass $\mathrm{log(M_{*}/M_{\odot})} > 10 $. This is intended to select SNe in older, massive, and passive hosts. Such a selection reduces our sample sizes to 48 (27 per cent removed) SNe Ia located within clusters and 516 (49 per cent removed) SNe Ia located in the field.

\begin{table*}
\caption{The number of SNe that are removed from our sample at each stage of selection.}
\begin{threeparttable}
\begin{tabular}{|c|c|c|}
\hline
Selection & Number remaining & Number removed \\
\hline
\begin{tabular}[c]{@{}l@{}}SNe pre-light curve cuts \\ \citep[see][]{Moller2022}\end{tabular} & 2802  & \\
SALT3 \& $P$(Ia) selection  & 1306  &     1496    \\
Redshift selection ($0.1 \leq z \leq 0.7$)       & 1154       &    152         \\
`Exclusion zone' cut \tnote{a}                &   1090      &    64          \\
\hline
Cluster SNe                             & 66         &  \\
Field SNe                             & 1024         & \\
\hline
\end{tabular}
\label{cuts}
\begin{tablenotes}
    \item[a] SNe within 1.5--2.5\,Mpc of a given cluster that also match the $pq$ limits described in Section~\ref{sec:Determining Supernovae Location} are excluded from the sample.
\end{tablenotes}
\end{threeparttable}
\end{table*}

\section{SN Ia properties in field and cluster galaxies}
\label{sec:Initial Results after Data Split}

Having defined our cluster and field SN Ia samples, we next compare their light curve properties, $c$ and $x_1$, as well as their host galaxy stellar masses. SALT3 $x_1$ is a measure of how quickly a SN's light curve evolves, with faster evolving events having lower values of $x_1$. The SALT3 $c$ of a SN is how red or blue the event is and encapsulates both intrinsic SN colour and reddening by host galaxy dust. SN $c$ and $x_1$ are empirically related to luminosity, via the linear \lq bluer--brighter\rq\ relationship and the \lq faster-fainter relation \rq.

\subsection{SN Ia light-curve width}
\label{sec:x1}
\begin{figure*}
    \centering
	\includegraphics[width=\columnwidth]{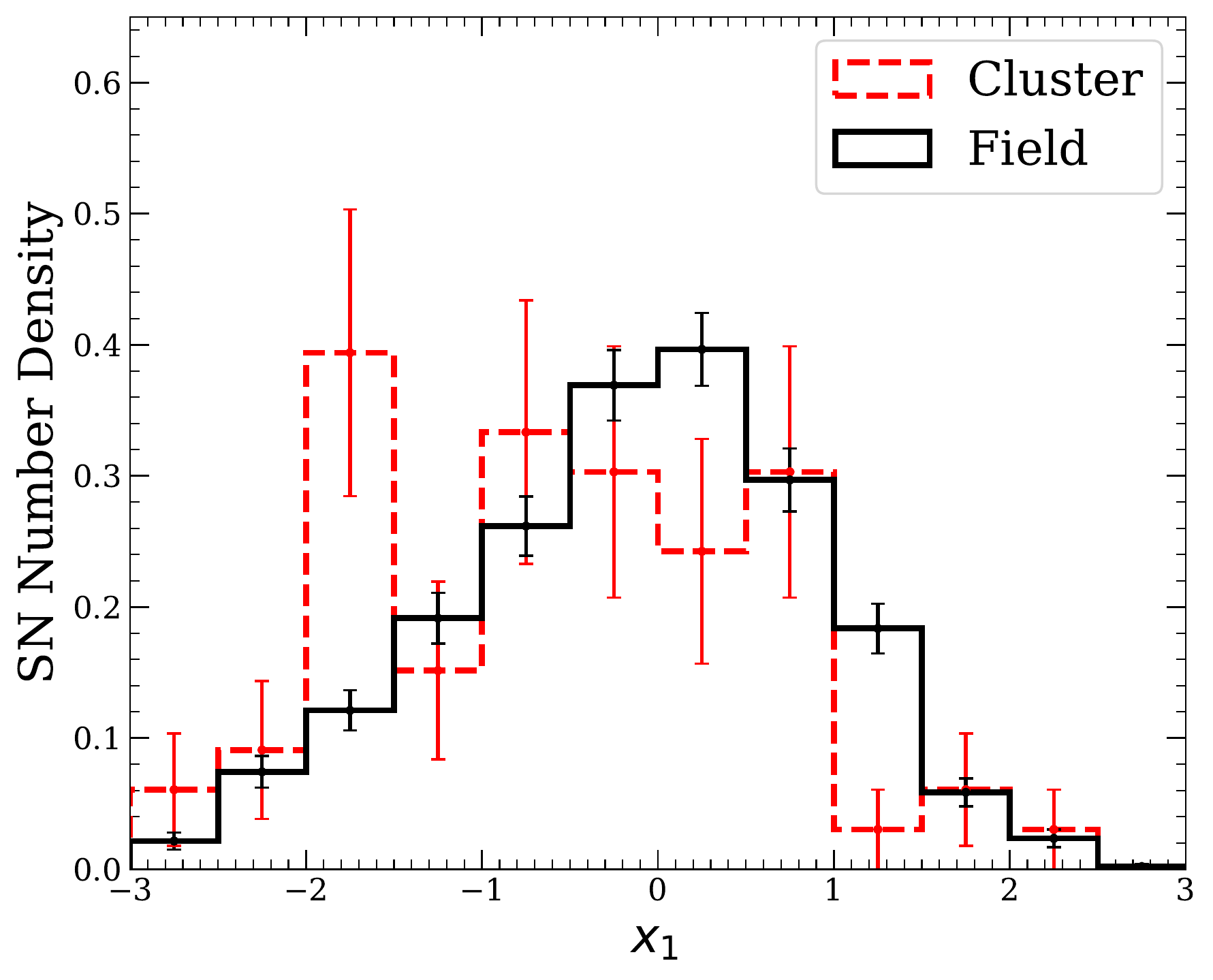}
	\includegraphics[width=\columnwidth]{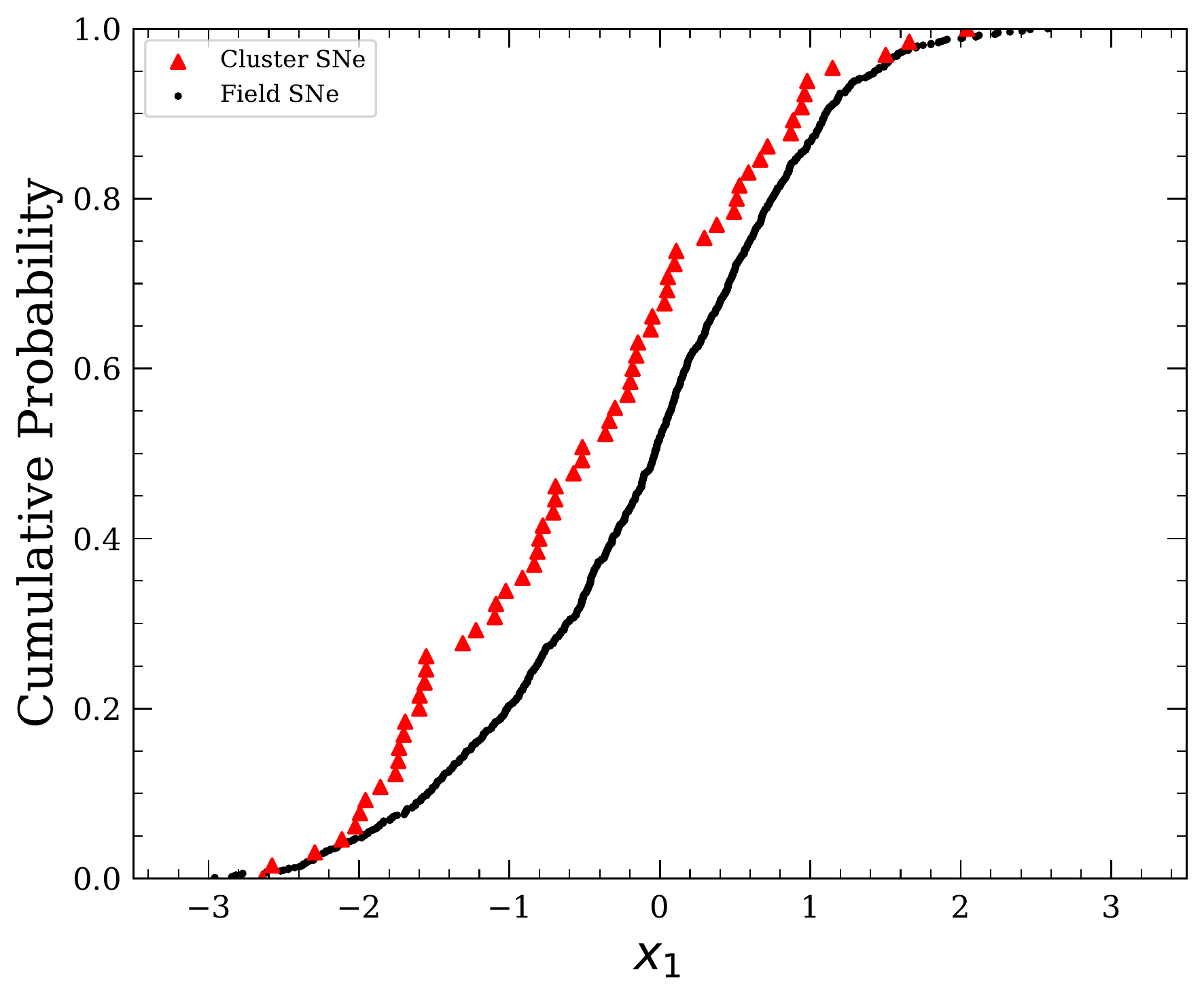}
    \includegraphics[width=\columnwidth]{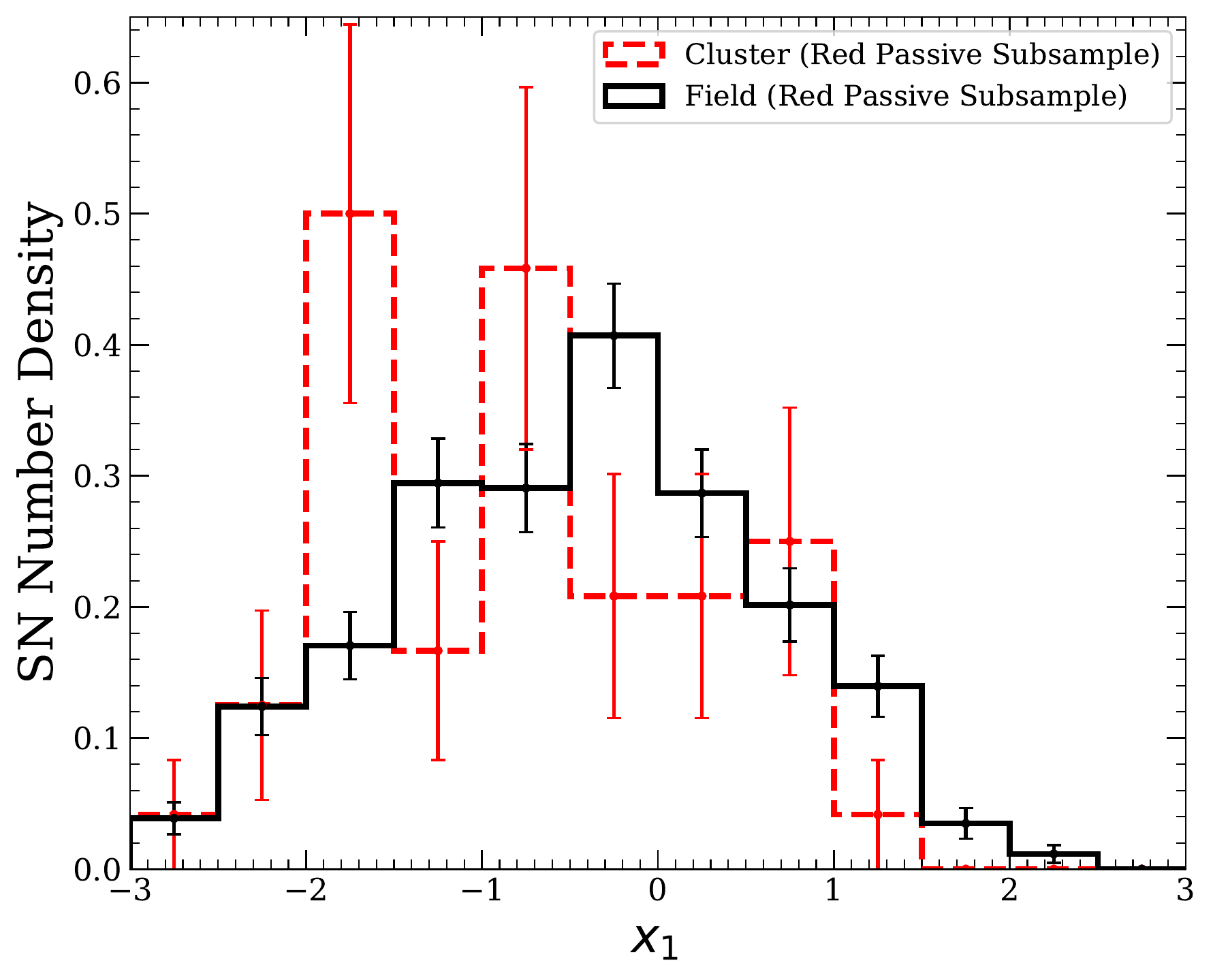}
    \includegraphics[width=\columnwidth]{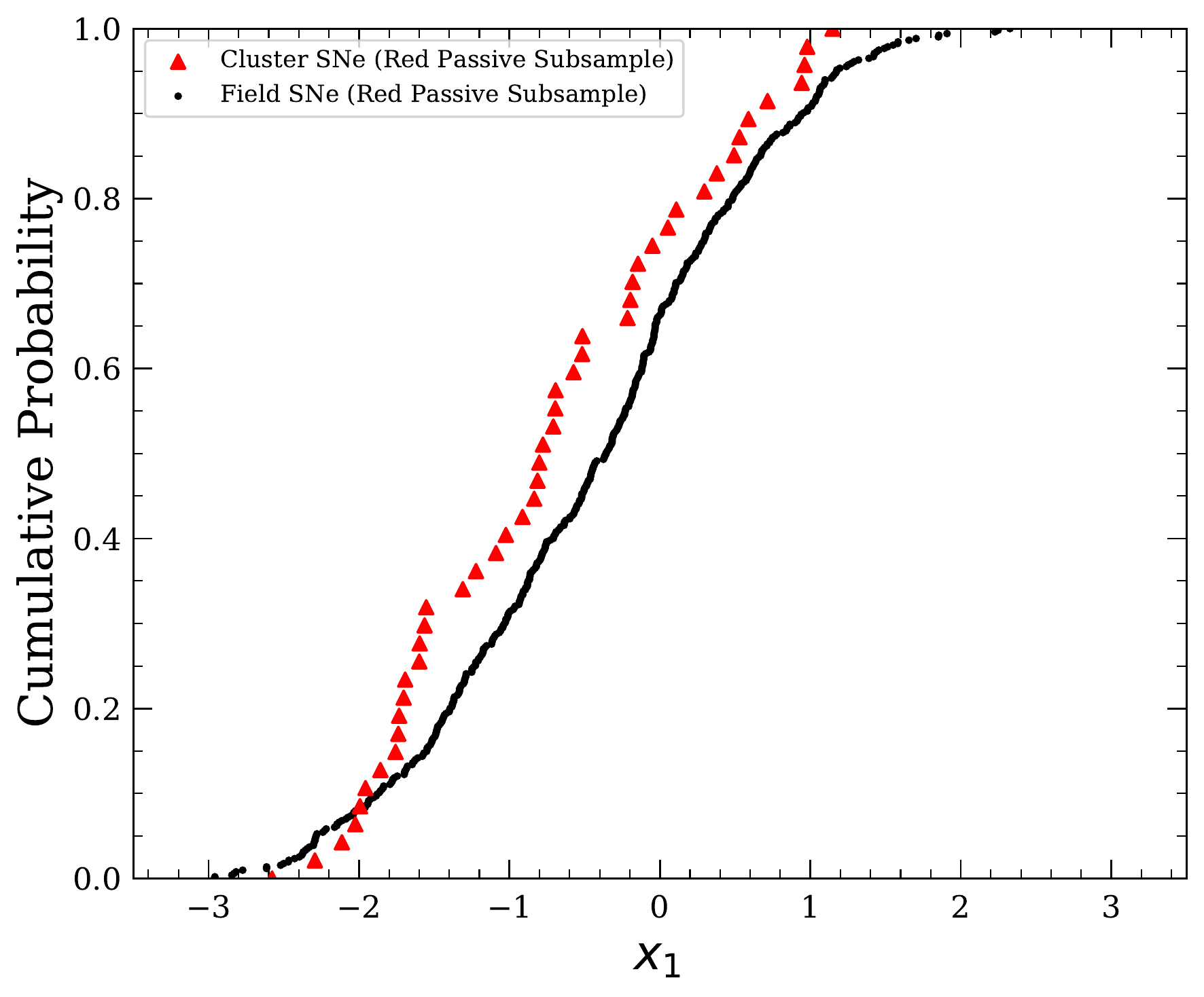}
	\caption{SN Ia $x_1$ distributions for events located in clusters versus those in the field. Upper Left: $x_1$ values in bins of width 0.5. Upper Right: CDFs of the $x_1$ data. A K-S test measures a $0.023$ probability that the distributions are drawn from the same parent distribution.
    In the lower plots, both samples have been selected to be only red and massive host galaxies (i.e., $U-R>1$ and $\mathrm{\log(M_{*}/M_{\odot})} \geq 10$) as described in Section~\ref{sec:Determining Supernovae Location}. A K-S test measures a $0.068$ probability that the distributions are drawn from the same parent distribution.}
    \label{fig:x1}
\end{figure*}

The $x_1$ distributions for cluster and field SNe, together with the cumulative distributions, are shown in the top two plots of Fig.~\ref{fig:x1}. The cluster distributions are shifted slightly to more negative $x_1$ values when compared to the field. A two-sided Kolmogorov–Smirnov (K-S) test returns a $p$-value of $0.023$, indicating that the two distributions are not drawn from the same parent distribution with a 97.7 per cent confidence level. This is a tentative confirmation of what we might expect to observe: cluster galaxies are typically more massive and passive than those in the field, and these galaxies typically host fainter, faster SNe Ia than galaxies with stronger star formation \citep{Hamuy1995}.

Previous results have found more significant differences in the $x_1$ parameter between cluster and field samples \citep{Xavier}. Their result however uses different light curve quality cuts. Furthermore they only consider rich galaxy clusters, while we make no such distinction.

The same analysis performed on our sub-sample of cluster and field SNe Ia in older, massive and passive galaxies is shown in the lower panels of Fig.~\ref{fig:x1}. The $x_{1}$ distribution is still shifted to the more negative values, but the significance is reduced with a K-S test $p$-value of 0.068. 

\subsection{SN Ia colour}
\label{sec:colour}

\begin{figure*}
    \centering
	\includegraphics[width=\columnwidth]{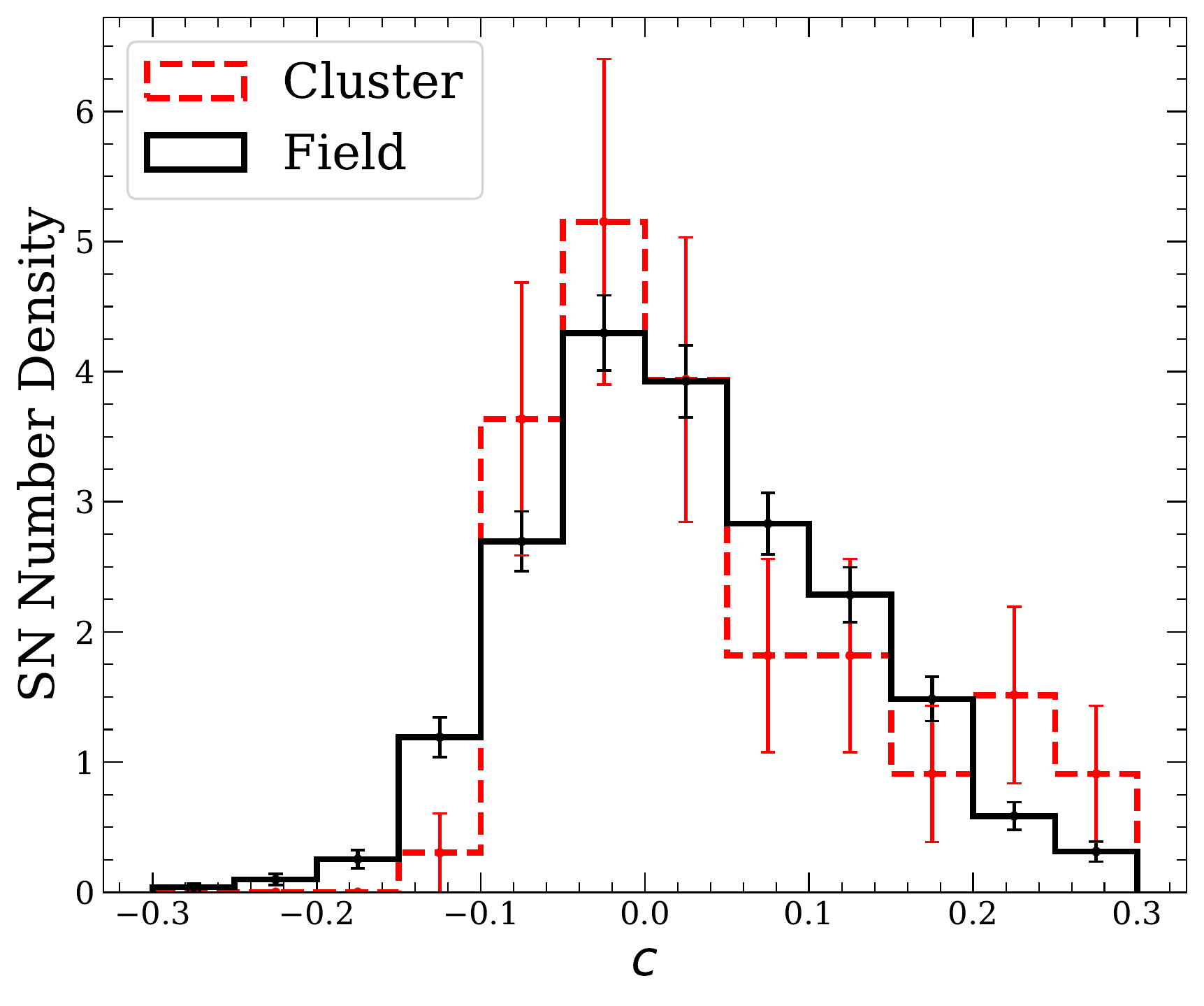}
	\includegraphics[width=\columnwidth]{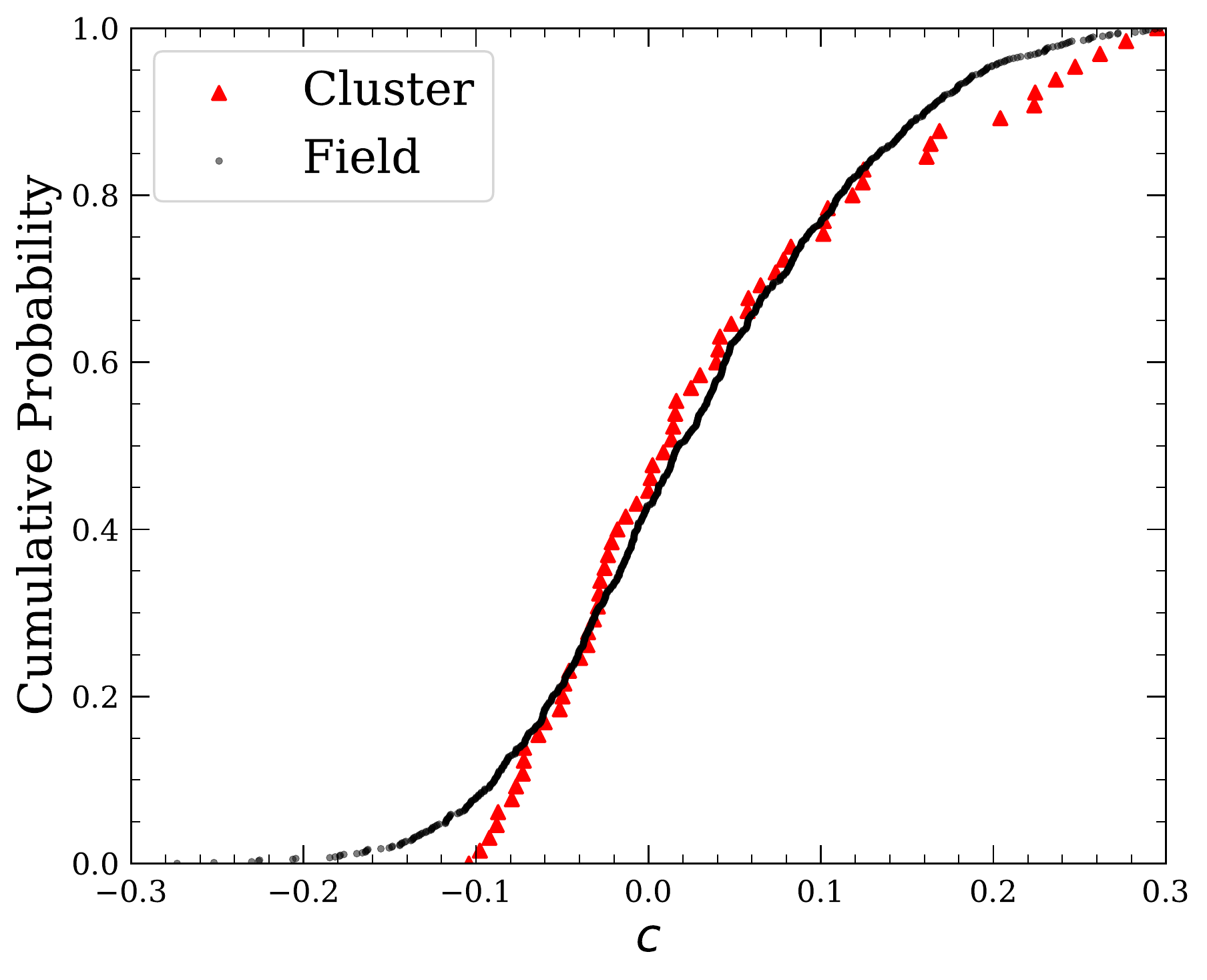}
    \caption{SN Ia colour distributions for events found in clusters and those found in the field. Left: Histograms binned in steps of 0.05. Right: CDFs of the colour distributions. A two sided K-S test gives a $p$-value of $0.801$, indicating no statistical evidence that the CDFs are from different parent distributions.} 
    \label{fig:c}
\end{figure*}

The SN colour distributions and cumulative distributions are shown in Fig.~\ref{fig:c}. The field and cluster distributions are consistent, with a K-S test $p$-value of $0.801$. Comparing the samples after the $U-R$ and host mass selection gives a similar $p$-value of $0.827$. Thus, we see no evidence in our sample for SN colour distributions that differ between cluster environments and the field. 

\subsection{SN Ia host galaxy stellar masses}
\label{sec:HostMass}

\begin{figure*}
    \centering
	\includegraphics[width=\columnwidth]{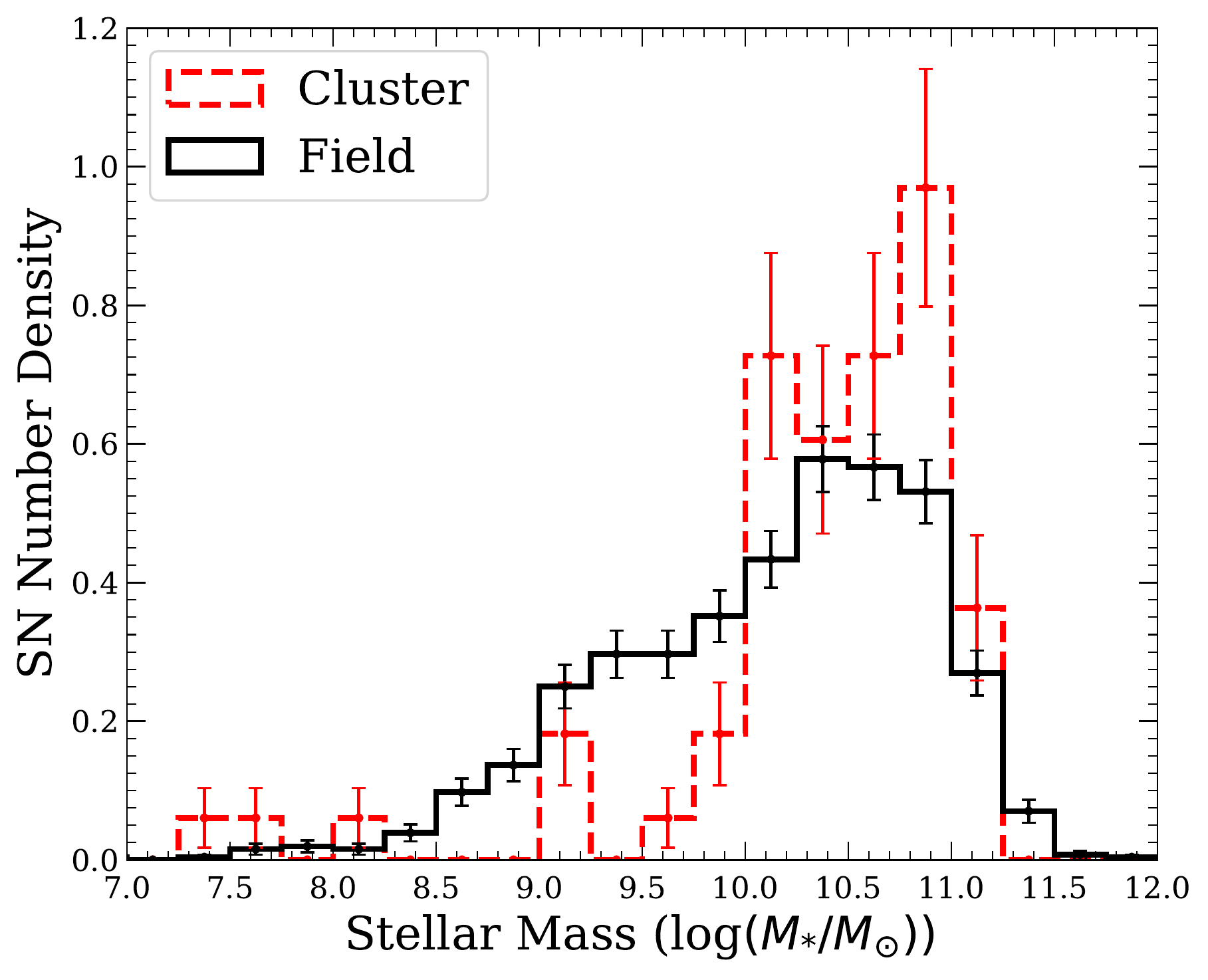}
	\includegraphics[width=\columnwidth]{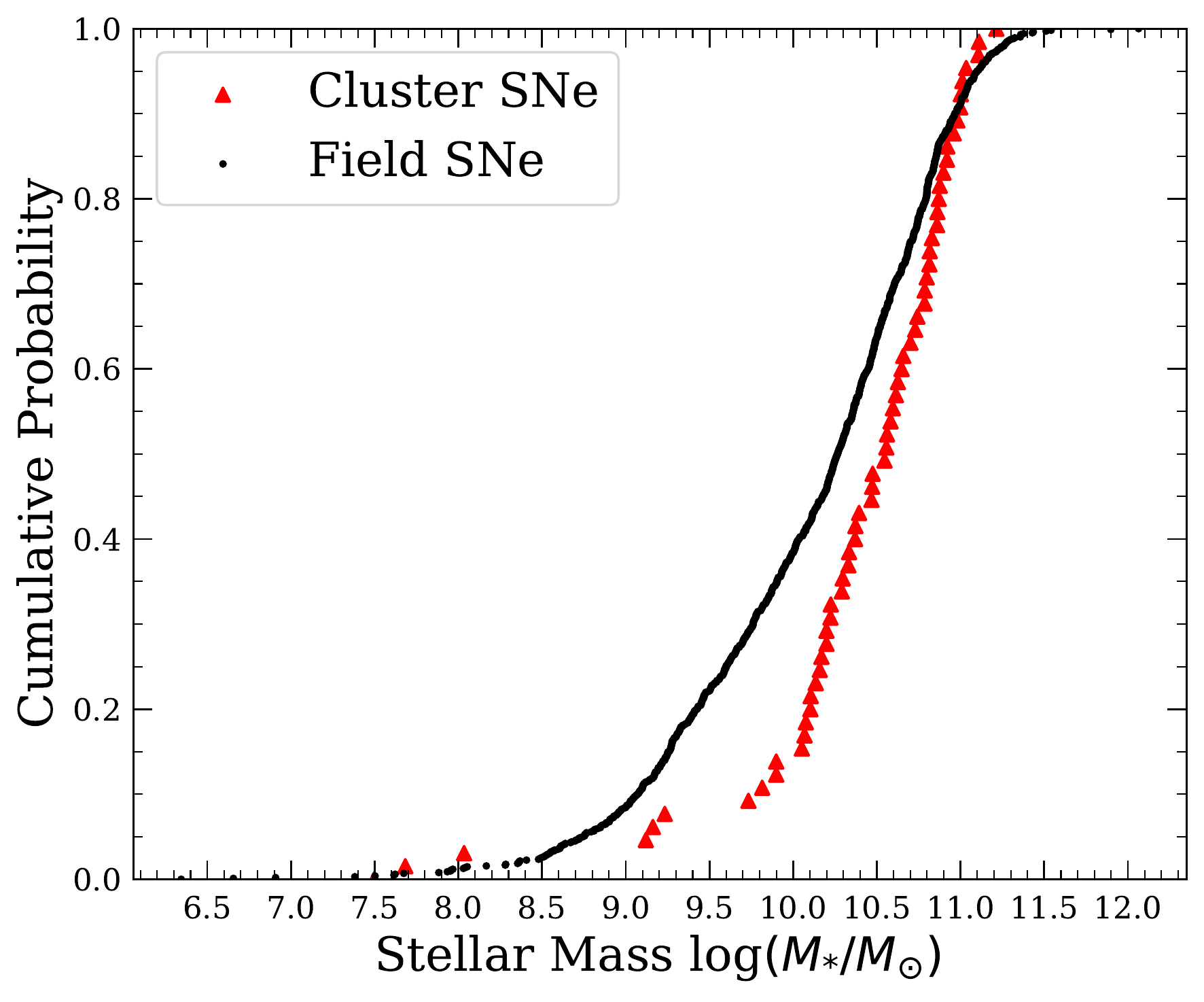}
    \caption{Host galaxy stellar mass distributions for SNe Ia occurring within clusters and in the field. Left: Cluster and field data binned in steps 0.25. There is a lack of cluster hosts with stellar masses between $9 \leq \mathrm{log(M_{*}/M_{\odot})} \leq 10$, skewing the distribution to the more massive end. Right: CDFs of the host stellar masses. A two-sided K-S test performed on the CDFs returned a $p$-value of $0.0006$, indicating strong evidence that the two distributions are drawn from different base distributions.}
    \label{fig:hmHist}
\end{figure*}

We show the distributions and cumulative distributions of SN Ia host galaxy stellar mass $M_*$ in Fig.~\ref{fig:hmHist}. As expected, there is a deficit of hosts with stellar masses of $\log(M_{*}/M_{\odot})\leq 10$ in clusters, with a K-S $p$ value of $0.0006$, i.e., the two distributions are drawn from different parent distributions with a significance of 3.6\,$\sigma$. As expected after selecting host mass and $U-R$ to probe similar galaxies, a two sided K-S test returns a $p$ value of 0.715, meaning statistically the two populations are drawn from the same parent distribution, reassuring us that our host mass and colour cut successfully facilitates a fair comparison of the SNe in these galaxies.

Possible explanations for the difference in stellar mass distributions for field and cluster SN host galaxies include: a different stellar mass function of cluster and field galaxies; a different rate of SNe Ia as a function of stellar mass in cluster and field galaxies, or a combination of the two. Such SN rate differences could be caused by the difference in age and star-formation activity between cluster and field galaxies. We examine these possibilities in the next section. 
 
\begin{figure}
    \centering
    \includegraphics[width=\columnwidth]{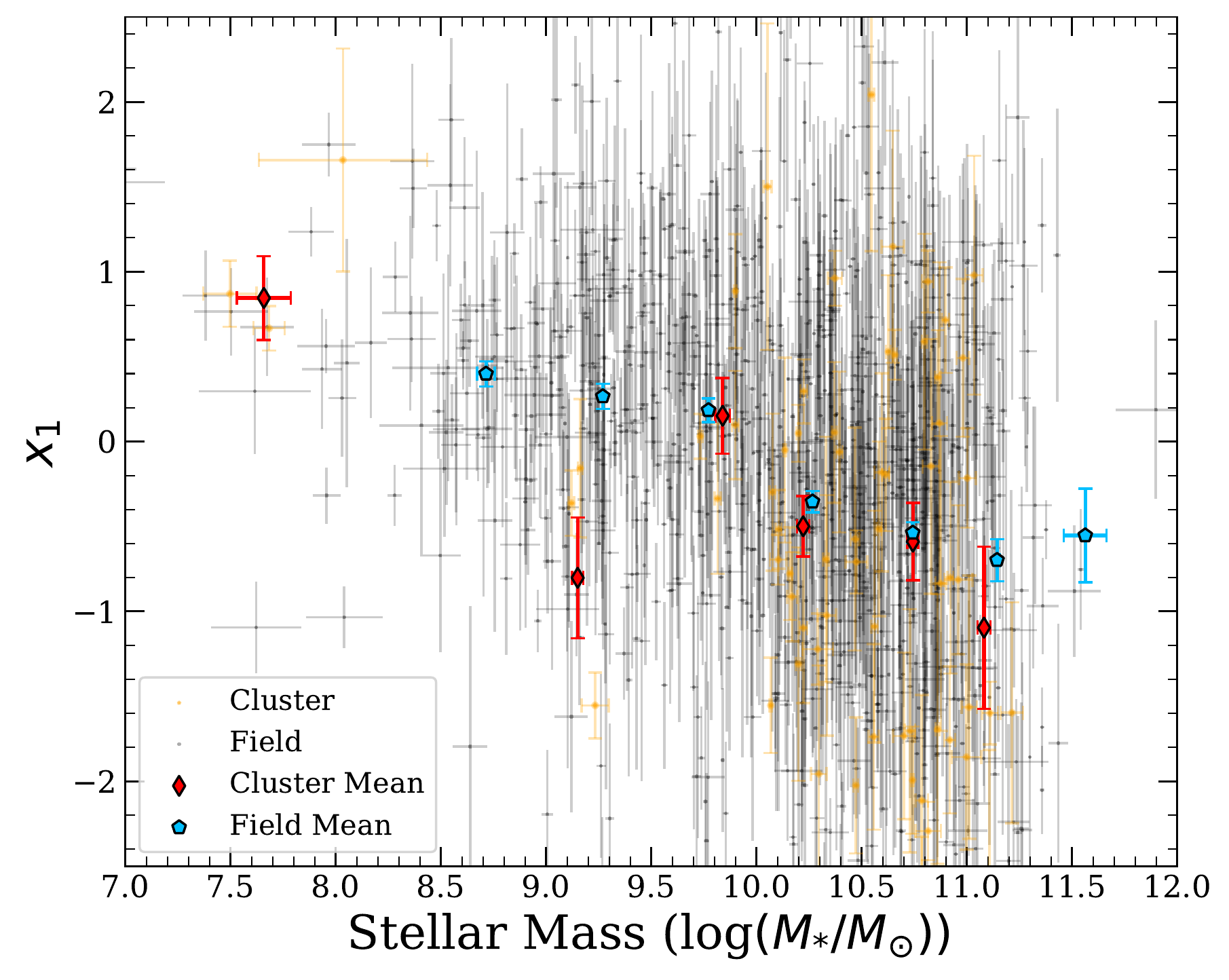}
    \caption{SN Ia $x_1$ versus host galaxy stellar mass for SNe Ia in clusters and the field, shown both as individual points (yellow stars/grey crosses) and as weighted means (red diamonds/blue pentagons). We recover the expected trend of higher mass hosts containing SNe Ia with smaller values of $x_1$. } 
    \label{fig:x1massmean}
\end{figure}

Fig.~\ref{fig:x1massmean} shows the SN Ia $x_1$ versus host stellar mass, where we recover the expected relationship: the average SN $x_1$ across both field and clusters is smaller in more massive hosts than in less massive hosts. In around half the bins, cluster SNe have smaller $x_1$ than their field counterparts, but the difference is  typically consistent within uncertainties in any one bin. 

\subsection{The Effect of Progenitor Age}
\label{sec:Further Investigation}

Galaxies in clusters have been found to be older than their similar-mass field counterparts \citep{Saracco2017}, with much of their stellar populations formed at $z > 2$ \citep{Guglielmo_2015}. Furthermore, stars near the centre of galaxies, regions with lower specific star formation rate, tend to be older \citep{Zheng2017}. 

For SNe Ia, \cite{Ivanov2000} and \cite{Lluis2012} find that SNe Ia in the centres of galaxies are fainter, and \cite{Howell2001} find that older progenitors lead to fainter SNe Ia. \cite{Rigault2020} provides an updated analysis, showing that $x_1$ correlates with specific star formation rate measured within a projected distance of 1\,kpc from each SN location (local sSFR, or LsSFR), and therefore progenitor age. However, the discovery rate of SNe in the centres of galaxies is relatively low \citep{Shaw1979}: regions of high surface brightness lead to smaller signal-to-noise in the SN detections, reducing the detection efficiency for SNe located near galaxy centres \citep{Kessler_2015}.

\begin{figure}
    \centering
	\includegraphics[width=\columnwidth]{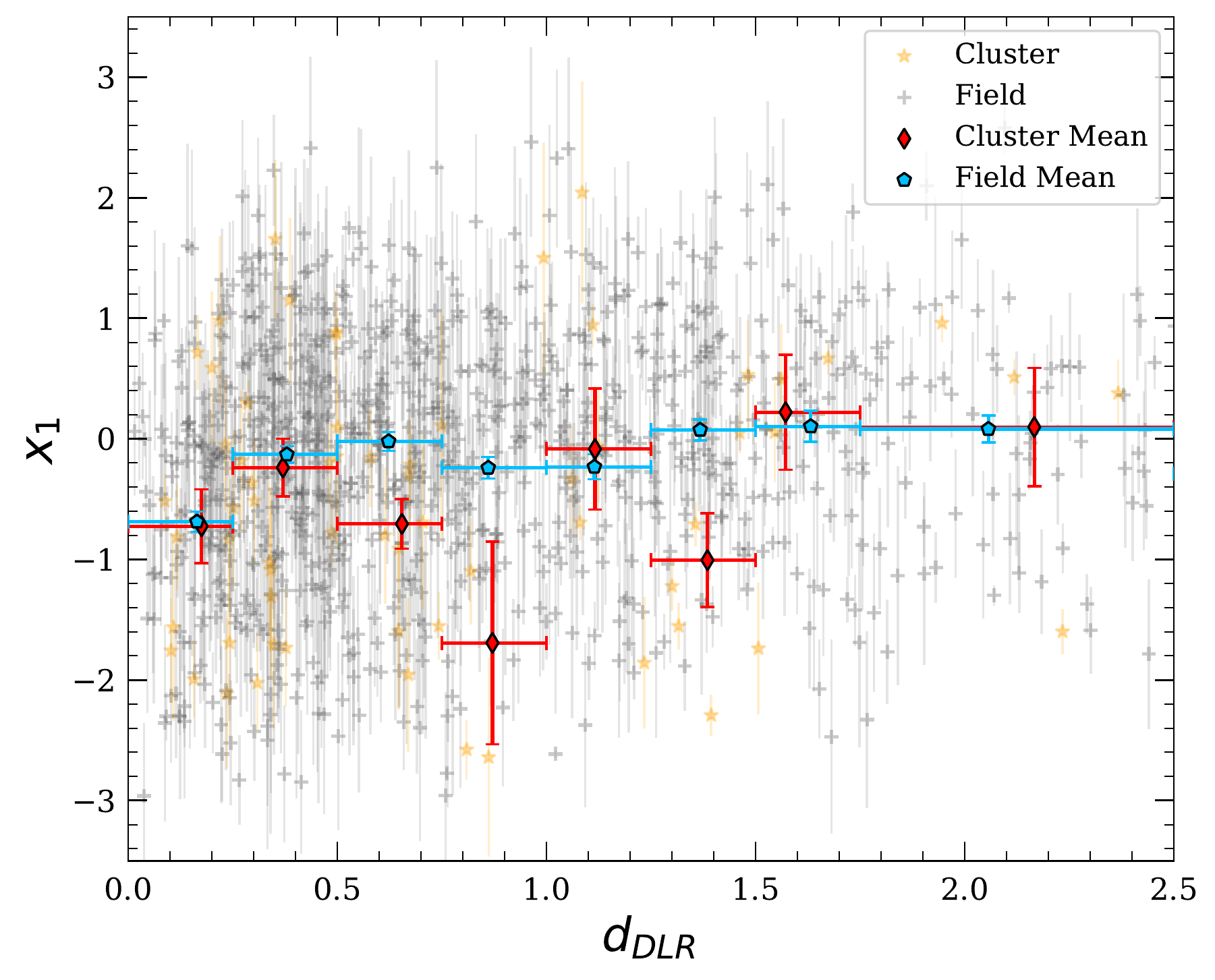}
    \caption{SN $x_1$ vs fractional host galaxy distance ($d_\mathrm{DLR}$) for SNe Ia in clusters and the field, with the mean values for $x_1$ within $d_\mathrm{DLR}$ bins of width 0.25 plotted. Due to few SNe Ia at these distances, we use one bin between $1.75 \leq d_\mathrm{DLR} \leq 2.5$. } 
    \label{fig:DLRx1mean}
\end{figure}

In Fig.~\ref{fig:DLRx1mean}, we show $x_1$ versus $d_\mathrm{DLR}$, a measurement of the effective distance of the SN from the galaxy centre \citep{Sullivan2006}. With the exception of the lowest $d_\mathrm{DLR}$ bin, in the field there is no trend between $x_1$ and $d_\mathrm{DLR}$; the weighted mean values for the cluster sample are broadly consistent with those of the field sample.
We confirm this visual lack of a trend by fitting a straight line to the data, which has a gradient consistent with zero. 

In the lowest $d_\mathrm{DLR}$ bin, there is a decrease in the error-weighted mean $x_1$ for SNe in both field and cluster galaxies (i.e., intrinsically fainter SNe Ia are preferentially located in this regions). The SNe in this $d_\mathrm{DLR}$ bin are closest to the host galaxy centres, where increased surface brightness would bias \textit{against} detecting fast, faint SNe. Similarly, increased dust extinction in these regions is unlikely to drive the absence of the intrinsically brighter / larger $x_1$ events. Thus, it is likely that the age gradient present in star-forming galaxies, strongest for $d_\mathrm{DLR} < 0.5$ \citep{Gonzalez-Delgado2015, Ibarra-Medel2016}, drives the effect that we see.

\section{Supernova Rates in Clusters and the Field}
\label{sec:Rates}

In Section \ref{sec:HostMass} we showed that there is a differing distribution of host galaxy stellar masses between SNe Ia within clusters and those in the field. In this section, we measure the  rate of SNe Ia per unit stellar mass -- the mass-normalised SN Ia rate, or the specific SN Ia rate -- as a function of the stellar mass of their host galaxies for the two samples.

\subsection{Calculating the SN Ia rate per unit stellar mass}

The rate of SNe Ia per unit stellar mass is calculated from the number of SNe Ia ($N_{\mathrm{SNe}}$) detected per unit time, divided by the total surveyed stellar mass. This rate can be further calculated as a function of stellar mass by repeating the calculation, but segregating events into bins based on the stellar mass of their host galaxies. 

The total amount of stellar mass within our two samples is calculated as follows. For the field, we estimate the total stellar mass by multiplying the ZFOURGE/CANDELS stellar mass function measured over $0.2 < z < 0.5$ \citep{ZFourge}  by the volume surveyed by the DES-SN survey over $0.1<z<0.7$. The ZFOURGE SMF, $\phi(M){dM}$, is described by the double Schechter function \citep{Schecter1976}
\begin{equation}
\begin{split}
\phi(M){dM} &= \phi_{1}({M}){dM} + \phi_{2}({M}){dM}\\
 &= \ln(10)\mathrm{e}^{-10^{(M-M^{*})}}10^{(M-M^{*})}\\
&\times[\phi_{1}^{*}10^{(M-M^{*})\alpha_{1}} + \phi_{2}^{*}10^{(M-M^{*})\alpha_{2}}]dM,
\label{SMF}
\end{split}
\end{equation}
where $M=\log(M/\mathrm{M}_{\odot})$, ($\alpha_{1}$, $\alpha_{2}$) are the slopes and ($\phi_{1}^{*}$, $\phi^{*}_{2}$) are the normalisations of the two Schechter functions, and $M^{*}$  is the characteristic mass. The product of the SMF and the field volume results in the galaxy numbers as a function of stellar mass that DES surveys, from which the total stellar mass in each mass bin can be calculated. 

For the cluster SN Ia sample, instead of calculating the volume encompassed by our clusters (which is uncertain) and multiplying by a cluster SMF, we instead use the relation between a cluster's richness, $\lambda$ and its total stellar mass, $M_\star$, i.e., 
\begin{equation}
\ln\left(\frac{M_\star}{\tilde{M}_\star}\right)=  \pi_{M_\star|\lambda} +\alpha_{M_\star|\lambda}\ln\left(\frac{\lambda}{\tilde{\lambda}}\right),
\label{eq:richnessmass}
\end{equation}
where $\tilde{M}_\star$ is the median ${M}_\star$ of the sample, and $\tilde{\lambda}=40$, $\tilde{\mathrm{z}}=0.35$ are the median richness and redshift used in \citet{McClintock2019}. Values for $\pi_{M_\star|\lambda}$, $\alpha_{M_\star|\lambda}$ and $\tilde{M}_\star$ are taken from Palmese et al. \textit{in prep} and \citet{Palmese2020}, which measured the stellar-to-halo mass relation for DES redMaPPer clusters.

This gives the overall mass for the cluster sample, which we separate into mass bins by using a SMF for DES clusters presented in Palmese et al. \textit{in prep}. However, this SMF is only valid for hosts with $\log(M_{*}/\mathrm{M}_{\odot}) \geq 10$. As such, we discard cluster SNe in hosts less massive than this limit for this rate analysis. Both field and cluster SMFs assume a Chabrier initial mass function (IMF). In SN Ia rate analyses, a Kroupa IMF is often used. As such we shift our IMFs to a Kroupa IMF, using Eq. 2 in \citet{Speagle_2014} which amounts to a difference of 0.01 dex.

$N_{\mathrm{SNe}}$ for both our samples is calculated as detailed in Section~\ref{sec:Determining Supernovae Location}. We account for time dilation, and the efficiency of the DES-SN survey in both detecting SNe and in measuring the redshift of the SN host galaxy using host galaxy spectroscopy. We do this following a standard \lq efficiency\rq\ method \citep[e.g.,][]{Perrett2012,Wiseman_2021}: we compute $\eta_{\mathrm{SN,i}}$, the detection efficiency of the $i$th SN, as 
\begin{equation}
    \eta_{\mathrm{SN,i}} = \eta_{\mathrm{SN,i}}(F_{i},z_{i},t_{0,i},x_{1,i},c_{i}) \times \epsilon_{z_{\mathrm{spec}}}(m^{\mathrm{host}}_{r,i}),
\end{equation}
where $\eta_{SN,i}(F_{i},z_{i},t_{0,i},x_{1,i},c_{i})$ is the SN detection efficiency of the $i$th SN exploding in field $F$, at time $t_0$ and redshift $z$, with stretch $x_1$ and colour $c$. This detection efficiency is estimated by simulating $1.1 \times10^6$ SNe Ia using \textsc{snana} and running a simulation of the DES-SN survey, and is fully described in \citet{Wiseman_2021}.

To obtain the detection efficiency, we  divide the number of SNe detected by the simulation of DES-SN (i.e., that pass the light-curve selection described in \ref{SNe Data}) by the total number of simulated SNe Ia. $\epsilon_{z_{\mathrm{spec}}}(m^{\mathrm{host}}_{r,i})$ is the efficiency of obtaining a spectroscopic redshift for our SN hosts as a function of $r$-band apparent magnitude, and has been modelled by \citet{vincenzi2021}. In the rate calculation, each SN is then weighted by the factor $1/\eta_{\mathrm{SN,i}}$.

To test the reliability of our detection efficiencies, we calculate a simple average volumetric rate ($\mathrm{SNR_{Ia}}$) for our field sample and compare to other analyses. We take the efficiency-corrected number of field SNe and divide by the co-moving volume within our redshift range. We calculate $\mathrm{SNR_{Ia}}(\langle z \rangle = 0.55) = 0.406 \times 10^{-4}\: \mathrm{SNe}\: \mathrm{yr^{-1}}\: \mathrm{Mpc^{-3}}$, consistent with \citet{Neill2006} who found $\mathrm{SNR_{Ia}}(\langle z \rangle = 0.47) = [0.42^{+0.13}_{-0.09}~(\mathrm{syst.}) \pm 0.06 ~(\mathrm{stat.})]\times 10^{-4}\: \mathrm{SNe}\: \mathrm{yr^{-1}}\: \mathrm{Mpc^{-3}}$ and \citet{Perrett2012} with $\mathrm{SNR_{Ia}}( 0.5< z < 0.6) = [0.48^{+0.06 + 0.04}_{-0.06 - 0.05}]\times 10^{-4}\: \mathrm{SNe}\: \mathrm{yr^{-1}}\: \mathrm{Mpc^{-3}}$. Our measurement is also within the 1$\sigma$ uncertainties of the power-law fit to the evolution of the $\mathrm{SNR_{Ia}}$ with redshift \citep{Frohmaier2019}. We have not performed a full uncertainty analysis on our $\mathrm{SNR_{Ia}}$ measurement as this is not the focus of this paper, and is deferred to future work.

\begin{figure}
    \centering
	\includegraphics[width=\columnwidth]{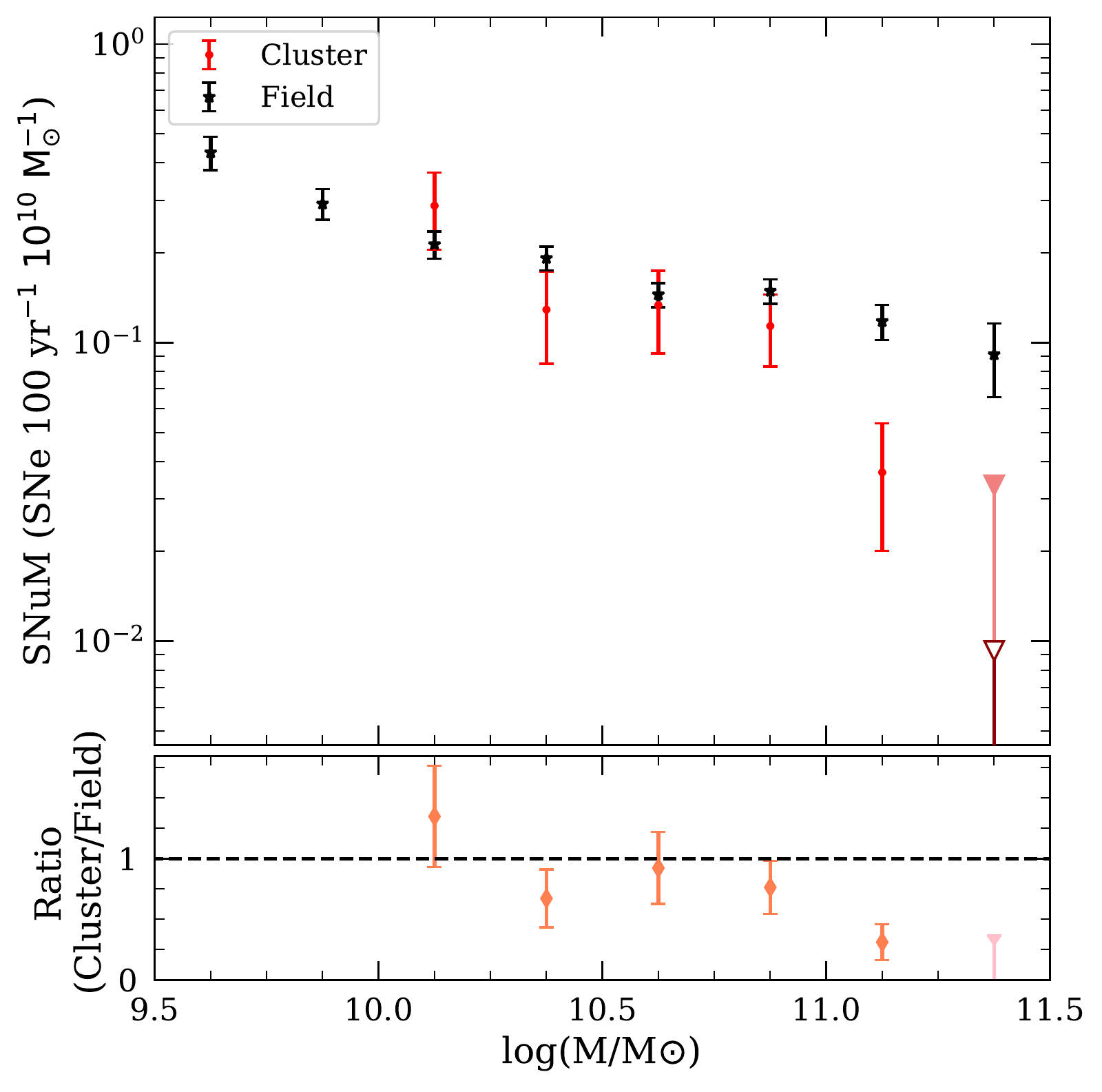}
    \caption{The number of SNe per $10^{10}~\mathrm{M}_{\odot}$ per century, as a function of stellar mass of the host galaxy for cluster SNe and field SNe. The final cluster data points are the 1$\sigma$ (dark red) and 3$\sigma$ (light pink) upper limits for a non detection from \citet{Gehrels1986}. The final ratio is the ratio of the 3$\sigma$ non-detection rate to the field.}
    \label{fig:rates}
\end{figure} 

We calculate the rate of SNe Ia per $10^{10} ~\mathrm{M}_{\odot}$ per century (also know as the SNuM) as a function of host galaxy stellar mass, shown in Fig.~\ref{fig:rates}. Within stellar mass bins that contain detected cluster SNe, the rate of SNe in cluster environments is broadly consistent with the field, with the respective rates mostly being  1$\sigma$ of each other. The largest outlier is the highest mass bin with cluster SNe detected, which sharply decreases compared to the field, with a difference of 3.6$\sigma$.
On average however, the rate of cluster SNe per mass is lower than the field, driven by the higher mass bin, with a weighted average ratio of the two of $0.594 \pm0.068$ between ($10\leq \log(M_{*}/\mathrm{M}_{\odot}) \leq 11.25$).

We do not detect any cluster SNe Ia in host galaxies above $\log(M_{*}/\mathrm{}M_{\odot}) = 11.25$, which is perhaps surprising. This lowers the overall cluster SNuM significantly, as a typical cluster SMF indicates galaxies with these stellar masses exist. For example, if the cluster rate from $11.25\leq \log(M_{*}/\mathrm{M}_{\odot}) \leq 11.5$ was equal to the field rate, $17.8\pm5.0$ SNe would have been expected over the 5-year DES-SN observing period given the total mass observed in that bin, and thus we would expect such objects to occur in our sample. However, cluster environments are also older when compared to the field, which (assuming a DTD that declines with time) also reduces the expected rate in these higher mass galaxies.

\begin{figure*}
    \centering
	\includegraphics[width=\columnwidth]{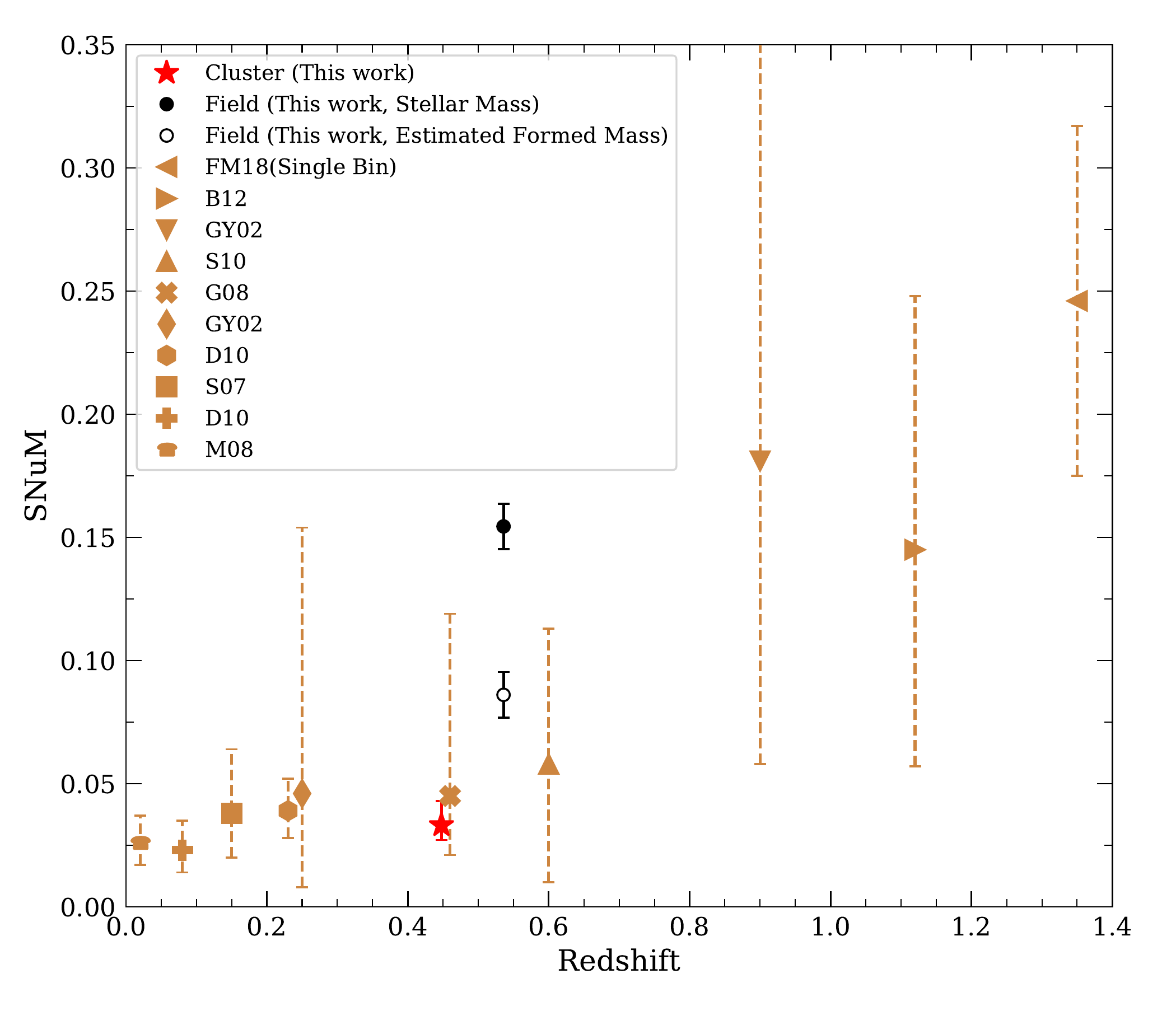}
 	\includegraphics[width=\columnwidth]{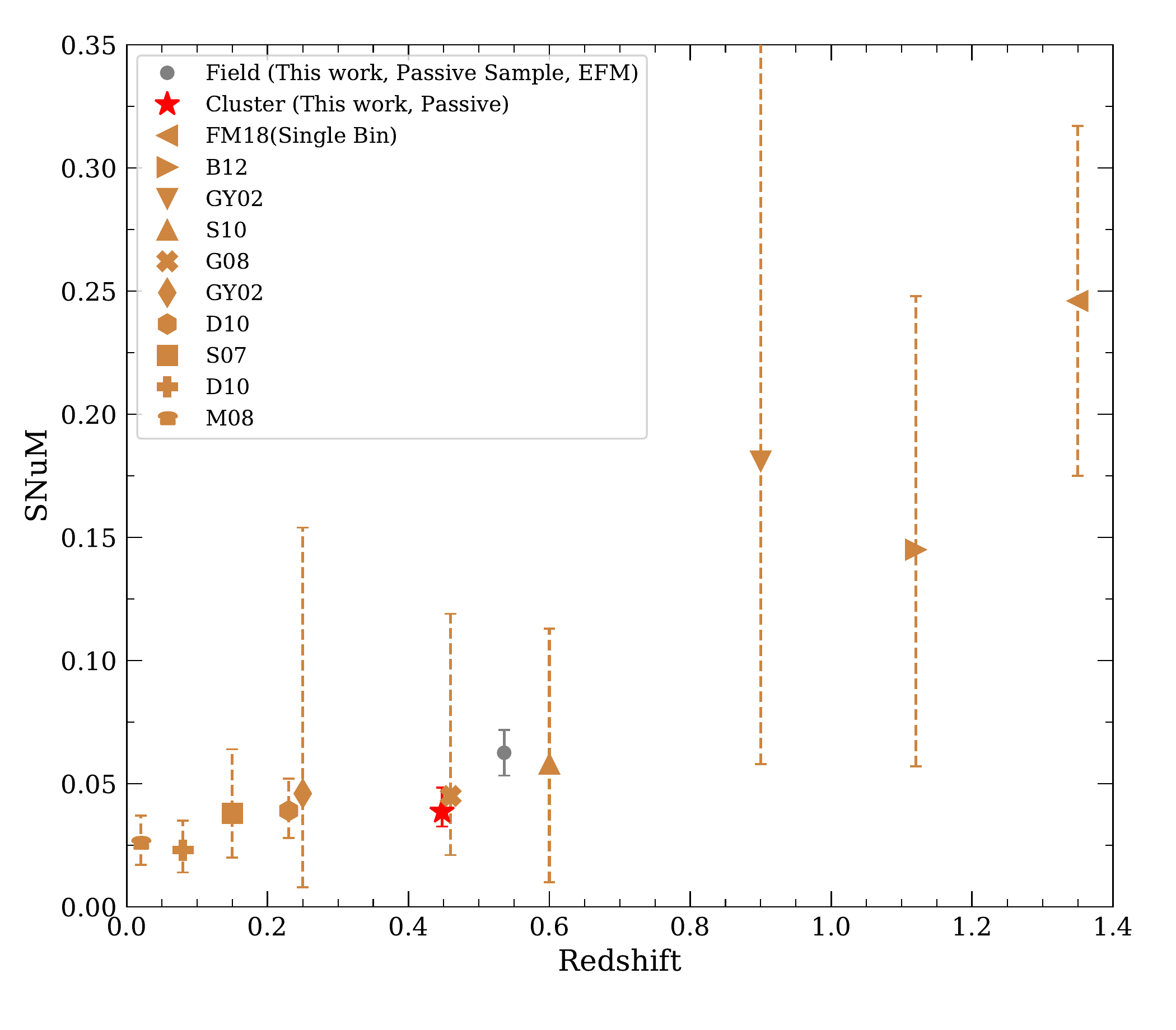}
    \caption{A comparison of our field and cluster SNuMs, in host galaxies with $\log(M_{*}/\mathrm{}M_{\odot}) \geq 10$,  compared to other literature examples. All cluster rates are in terms of formed mass, not stellar mass. The gold markers are rates measured in clusters from the literature. Left: The black filled point is the field rate per stellar mass, and the black circle is the estimated field rate per formed mass, assuming a similar SFH to cluster environments, the error of which should be treated as a lower limit. Right: A similar comparison, but instead both field and cluster have been limited to galaxies passing the UVJ cut described in Section \ref{sec:NIRdata}. Both cluster and field are presented in terms of formed mass. Errors on these passive rates should be treated as a lower limit. The redshift range of this work spans $0.1 \leq z \leq 0.7$. Data are in Table \ref{tab:rates}.} 
    \label{fig:ratecompare}
\end{figure*}

Taking into account the SMF above $\log(M_{*}/\mathrm{M}_{\odot})= 10$ for clusters and field, we calculate the integrated SNuM measurement for both environments. SNuM is known to be a strong function of galaxy properties \citep{Mannucci2005, Sullivan2006, Smith2012, Wiseman_2021}, and as such this host mass selection is key to making sure our results are a fair comparison, and do not include less massive field hosts, which have a higher SNuM then higher mass hosts. We find the integrated SNuM for clusters to be $0.0332\pm 0.0040 ^{+0.0082}_{-0.0044}$ at a efficiency weighted redshift of 0.44, while the field SNuM is $0.086\pm 0.0069 \pm 0.0062$ at a efficiency weighted redshift of 0.54. There exists an evolution of rate with redshift, which will account for some of the difference between our cluster and field rate.  We compare our measurements of total SNuM to measurements from the literature in Fig.~\ref{fig:ratecompare} and Table~\ref{tab:rates}.

We also calculate the integrated mass of the stars formed (formed mass) in our cluster hosts to compare the production efficiency of SNe Ia between the field and clusters, and for comparison to the literature. For this, for clusters we assume a single burst of star formation at $z=3$, and a constant metallicity of 0.02, using the P\'egase.3 \citep{Pegase3} spectral synthesis code as done in \citet{Freundlich2021}, and present our overall rate as a function of formed mass rather than stellar mass. After cutting low mass galaxies, we assume that the star formation history of the field hosts is the same as the cluster hosts. This allows us to make a simple estimate of the formed mass for the field galaxies, shown as the unfilled black circle in Fig.~\ref{fig:ratecompare}. 

Our cluster SNuM is the most precise in its redshift range, and is consistent with other measurements at lower and higher redshifts. However, we stress for these overall rates we have removed both cluster and field SNe that are located in low mass galaxies. This would lower our overall rate compared to literature rates. Additionally, due to the cluster SMF being only valid for use in galaxies $\log(M_{*}/\mathrm{}M_{\odot}) \geq 10$ we have assumed all cluster mass is made up of these galaxies. As we have found cluster SNe Ia hosted in less massive galaxies (Figs. \ref{fig:zcut}, \ref{fig:U_R_Host}, \ref{fig:hmHist}, \ref{fig:x1massmean}) we know this to be an over-simplification. As such we estimate the amount of mass below this limit in the cluster using the ZFOURGE/CANDELS passive stellar mass function (measured over $0.2 < z < 0.5$). We account for this uncertainty in the mass in the upper error on our cluster SNuM. Our field rate is also consistent with literature cluster rates at similar redshifts.

\begin{table*}
\centering
\caption{Overall SN Ia rates (with host mass cuts applied) from this work with comparisons from the literature, as recalculated by \citet{Freundlich2021}.}
\begin{threeparttable}

\begin{tabular}{|c|c|c|}
\hline
\multicolumn{1}{|l|}{} & \begin{tabular}[c]{@{}c@{}} Rate (SNuM) \end{tabular} & Average Redshift \\ \hline
Cluster SNuM  (This work)  &  $0.0332\pm 0.0040 ^{+0.0082}_{-0.0044}$\tnote{a}  & $0.44$\\ 
Field SNuM (This work)\tnote{b} &  $0.086\pm 0.0069 \pm 0.0062$\tnote{b}  & $0.54$\\ 
Field SNuM (Passive galaxies) &  $0.0625\pm 0.0069 \pm 0.0062$\tnote{c}  & $0.54$\\
\hline
\multicolumn{3}{c}{Cluster literature SN Ia rates} \\
\citet{Friedmann_2018}\tnote{d} & $0.246^{+0.071}_{-0.071} $& 1.12 \\
\citet{Barbary2012}  & $0.145^{+0.103}_{-0.088} $& 1.12 \\
\citet{Galyam2002}  & $0.181^{+0.241}_{-0.123} $& 0.9 \\
\citet{Sharon2010}  & $0.058^{+0.055}_{-0.048}$ & 0.6\\
\citet{Graham2008} & $0.045 ^{+0.074}_{-0.024} $& 0.46 \\
\citet{Galyam2002}  & $0.046^{+0.108}_{-0.038} $& 0.25 \\
\citet{Dilday_2010}  & $0.039^{+0.013}_{-0.011}$ & $0.23$ \\
\citet{Sharon2007}   & $0.038 ^{+0.026}_{0.018}$& 0.15 \\
\citet{Dilday_2010} & $0.023^{+0.012}_{-0.009}$& 0.08 \\
\citet{Mannucci2008} & $0.026^{+0.011}_{-0.009} $& 0.02 \\

\hline
\end{tabular}
\begin{tablenotes}
    \item[a] Statistical + systematic + SMF Correction
    \item[b] This is for a estimated formed mass, assuming a similar SFH as the cluster hosts. Errors presented are for a stellar mass based SNuM, and as such should be treated as a lower boundary.
    \item[c] Minimum error, see Section \ref{sec:NIRdata}.
    \item[d] Single redshift bin
\end{tablenotes}
\end{threeparttable}
\label{tab:rates}
\end{table*}

\subsection{Discussion of rates}

Our SN Ia SNuM in galaxy clusters is lower than that in the field, whether comparing overall values or only rates for SNe Ia in similar mass hosts. Fig. \ref{fig:rates} also shows that the ratio of the cluster rate compared to the field rate tentatively decreases with host stellar mass. We discuss the origin of this trend below.

\subsubsection{The effect of age on the SN Ia rate}

The cluster DTD has been measured to be normalised higher than the field DTD by many studies \citep{Moaz2017, Freundlich2021}. This enhanced DTD could be caused by an excess of white dwarfs (WD), due to a differing IMF, or an enhancement of binary systems within clusters compared to the field \citep{Friedmann_2018}. For example, the IMF may be non-universal \citep{VanDokkum2008, Romeel2008} and  depend on a galaxy's velocity dispersion, which may in turn lead to an excess of low-mass stars in the most massive galaxies \citep{Ferreras2013, FMateu2013}. This may lead to cluster galaxies' stellar populations containing a higher fraction of low-mass stars than the field, which, evolved over long enough times could lead to more WDs, causing a higher normalised DTD.

One may expect due to this increased DTD, to measure more SNe within clusters than in the field at the same redshift. However we find that the overall rate of SNe Ia per unit formed mass in massive galaxies within the field environments is higher than cluster environments.
Assuming a declining DTD, our lower rate in clusters would indicate that cluster galaxies of the same mass have older stellar populations than their field counterparts. This age dependent rate could negate this higher-normalised DTD, and explain the damped cluster rate.

It may be possible to probe the effect of an older stellar population by measuring the $U-R$ colour of SN host galaxies. $U-R$ can be used as a proxy for morphology, and is dependent on the star formation history of the galaxy being studied \citep{Lintott2008}. $U-R$ also correlates with galaxy age \citep{Wiseman2022}, with redder galaxies being older, and thus we can use $U-R$ to probe the galaxy age. 
We have shown the $U-R$ colours versus their stellar masses for our two samples in Fig.~\ref{fig:U_R_Host}. We see a slightly higher proportion of high mass, red galaxies within clusters compared to the field. The increasing $U-R$ with galaxy stellar mass, and the higher proportion of red galaxies in the cluster SNe hosts, may help explain the dampening in the rate as a function of mass in the cluster population. 

A possible cause of this older population is that star formation in clusters turned off earlier than in the field. The rate of galaxy mergers within cluster environments compared to the field may be enhanced \citep{Watson2019}, although the significance of this increase is debated, and evidence also exists for a comparable or lower cluster merger rate when comparing the field to the central cluster environment \citep{Delahaye2017}. Galaxy mergers can alter the gas content of the interacting galaxies, with some gas potentially being removed from one galaxy and taken by another, or lost to the intergalactic medium. Such a removal of gas would quench star formation within clusters, leading to older populations and a subsequent decrease in the rate.

Further investigation is needed into the effect of the IMF on WD production efficiency, and on galaxy mergers causing a differing age in similar mass galaxies in different environments. This would allow further constraints on the cluster and field DTD in order to see what effects this would have on SN production over Hubble Time. A precise DTD in galaxy clusters and massive field galaxies is deferred to a future work.

\subsubsection{The SN Ia rate in passive galaxies}
\label{sec:NIRdata}
We are able to estimate the effects of an age difference between the cluster and the field environments by isolating field galaxies that are passive, and thus should have older stellar populations. 
To calculate the SNuM in massive, passive field galaxies requires a passive field galaxy stellar mass function, which is provided by ZFOURGE. The ZFOURGE passive galaxies were identified using the rest-frame U, V, and J bands. 

To estimate the total number of massive field galaxies that are passive, we compute the fraction of those with NIR coverage that pass the \cite{ZFourge} UVJ cut and multiply it by the total number of galaxies. We then calculate the total passive formed mass using the passive SMF in \citet{ZFourge} and assuming a similar SFH to clusters, as above. We can then calculate the rate per century per $10^{10} ~\mathrm{M}_{\odot}$ formed, which we find to be $0.0625$ with the minimum error being $\pm 0.0069 ~\mathrm{(stat.)}\pm 0.0062~\mathrm{(syst.)}$. As we have made some oversimplifications we do not know the exact error on this rate. However, assuming similar or slightly enhanced errors to our other field SNuMs, this passive rate is comparable (different at a maximum of 1.7$\sigma$) to the passive cluster rate of $0.0386$ $\pm 0.0040~\mathrm{(stat.)} ~^{+0.0082}_{-0.0044}~\mathrm{(syst.)}$ (again the minimum error). Additionally, fitting a straight line to the literature values in Fig. \ref{fig:ratecompare} allows us to probe the evolution of the SNuM as a function of redshift. We find that such a rate evolution has a gradient of $\sim0.11\pm0.02$. Thus the average redshift difference of ~0.1 between our cluster and field samples would have a related rate difference of $\sim0.011$. Shifting the cluster rate to the field redshift using this evolution brings the maximum difference to $<1\sigma$.

Therefore it appears that while the stellar population in cluster host galaxies may be much older than those within overall field hosts, they are comparable in age or slightly older to stellar populations in passive field hosts. 

We however note that this passive rate contains many simplifications, and more wide-reaching NIR data would be needed to allow these comparisons to be performed in a more precise manner.

\section{Summary}
\label{sec:Conclusions}
Using the DES 5-year photometrically-confirmed type Ia supernova (SN Ia) sample, we have identified 66 SNe Ia that have occurred within redMaPPer clusters of galaxies, the largest high-redshift sample of cluster SNe Ia to date. We analysed and compared the light-curve  and environmental properties of this SN Ia sample to 1024 DES SNe Ia that  occurred in the field. We have also calculated the rate of SNe Ia per $10^{10} ~\mathrm{M}_{\odot}$ per century for the two samples, as a function of host galaxy stellar and formed mass. Our main findings can be summarised as follows.

\begin{itemize}
    
    \item We find a tentative indication that the light curve widths, $x_1$, of cluster SNe Ia are, on average, more negative (i.e., fainter and faster evolving) than their field counterparts. Although this is an expected result, as $x_1$ has a known  dependence on galaxy stellar mass and age, the evidence is not strong in this sample. When just comparing galaxies with similar host masses and colours, the significance drops further, perhaps implying that any differences we see are due to the lack of low-mass, young cluster hosts. We find that the colours of cluster SNe Ia statistically match those of the field, with very similar distributions.
    \item There is no clear relationship between a SN's $x_1$ and its location in its host. The exception is for the innermost SNe Ia, which have smaller values of $x_1$ for both cluster and field hosts.
    \item We calculate the rates of SNe Ia in cluster and field environments, and find them to be broadly consistent to within $1\sigma$. However, at higher masses this appears to change, with the rate of cluster SNe between ($11\leq \mathrm{log(M_{*}/M_{\odot})} \leq 11.25$) being lower than the field by $3.6\sigma$. Taking into account all mass bins with a detected cluster SNe, the weighted average ratio of cluster to field SNe rates is 0.594 $\pm$ 0.068, however this average ratio is heavily driven by the final cluster mass bin. 
    \item Integrating the overall rates of field and clusters within galaxies with $\mathrm{log(M_{*}/M_{\odot})} \geq 10.0$, we find the rate of SNe Ia within clusters as a fraction of formed mass to be $0.0332\pm 0.0040 ~\mathrm{(stat.)} ^{+0.0082}_{-0.0044} ~\mathrm{(syst.)}$ SNe 100 yr$^{-1}$ $10^{10}~\mathrm{M}_{\odot}^{-1}$, and the corresponding field rate to be $0.086\pm 0.0069~ \mathrm{(stat.)} \pm 0.0062 ~\mathrm{(syst.)}$  SNe 100 yr$^{-1}$ $10^{10}~\mathrm{M}_{\odot}^{-1}$. However, these measurements are at slightly differing redshifts, which will account for some of this difference and both are broadly consistent with other literature cluster rates. The measured decrease could be due to cluster galaxies being older, or more quenched than their field counterparts. Thus this declining rate within clusters compared to the field indicates that galaxies at a fixed mass are older in clusters than the field. We calculate that the rate in passive field galaxies is more comparable to the cluster rate, however a more complete dataset would be valuable in verifying this result.
\end{itemize}

\section*{Acknowledgements}
The authors wish to thank Kathryn Moser for their useful discussions and contributions to the paper.

The authors additionally wish to thank the anonymous reviewer for their detailed comments and support in improving this paper.

P.W. acknowledges support from the Science and Technology Facilities Council (STFC) grant ST/R000506/1.
L.G. acknowledges financial support from the Spanish Ministerio de Ciencia e Innovaci\'on (MCIN), the Agencia Estatal de Investigaci\'on (AEI) 10.13039/501100011033, and the European Social Fund (ESF) "Investing in your future" under the 2019 Ram\'on y Cajal program RYC2019-027683-I and the PID2020-115253GA-I00 HOSTFLOWS project, from Centro Superior de Investigaciones Cient\'ificas (CSIC) under the PIE project 20215AT016, and the program Unidad de Excelencia Mar\'ia de Maeztu CEX2020-001058-M.
L.K. thanks the UKRI Future Leaders Fellowship for support through the grant MR/T01881X/1.

Funding for the DES Projects has been provided by the U.S. Department of Energy, the U.S. National Science Foundation, the Ministry of Science and Education of Spain, 
the Science and Technology Facilities Council of the United Kingdom, the Higher Education Funding Council for England, the National Center for Supercomputing 
Applications at the University of Illinois at Urbana-Champaign, the Kavli Institute of Cosmological Physics at the University of Chicago, 
the Center for Cosmology and Astro-Particle Physics at the Ohio State University,
the Mitchell Institute for Fundamental Physics and Astronomy at Texas A\&M University, Financiadora de Estudos e Projetos, 
Funda{\c c}{\~a}o Carlos Chagas Filho de Amparo {\`a} Pesquisa do Estado do Rio de Janeiro, Conselho Nacional de Desenvolvimento Cient{\'i}fico e Tecnol{\'o}gico and 
the Minist{\'e}rio da Ci{\^e}ncia, Tecnologia e Inova{\c c}{\~a}o, the Deutsche Forschungsgemeinschaft and the Collaborating Institutions in the Dark Energy Survey. 

The Collaborating Institutions are Argonne National Laboratory, the University of California at Santa Cruz, the University of Cambridge, Centro de Investigaciones Energ{\'e}ticas, 
Medioambientales y Tecnol{\'o}gicas-Madrid, the University of Chicago, University College London, the DES-Brazil Consortium, the University of Edinburgh, 
the Eidgen{\"o}ssische Technische Hochschule (ETH) Z{\"u}rich, 
Fermi National Accelerator Laboratory, the University of Illinois at Urbana-Champaign, the Institut de Ci{\`e}ncies de l'Espai (IEEC/CSIC), 
the Institut de F{\'i}sica d'Altes Energies, Lawrence Berkeley National Laboratory, the Ludwig-Maximilians Universit{\"a}t M{\"u}nchen and the associated Excellence Cluster Universe, 
the University of Michigan, NSF's NOIRLab, the University of Nottingham, The Ohio State University, the University of Pennsylvania, the University of Portsmouth, 
SLAC National Accelerator Laboratory, Stanford University, the University of Sussex, Texas A\&M University, and the OzDES Membership Consortium.

Based in part on observations at Cerro Tololo Inter-American Observatory at NSF's NOIRLab (NOIRLab Prop. ID 2012B-0001; PI: J. Frieman), which is managed by the Association of Universities for Research in Astronomy (AURA) under a cooperative agreement with the National Science Foundation.

The DES data management system is supported by the National Science Foundation under Grant Numbers AST-1138766 and AST-1536171.
The DES participants from Spanish institutions are partially supported by MICINN under grants ESP2017-89838, PGC2018-094773, PGC2018-102021, SEV-2016-0588, SEV-2016-0597, and MDM-2015-0509, some of which include ERDF funds from the European Union. IFAE is partially funded by the CERCA program of the Generalitat de Catalunya.
Research leading to these results has received funding from the European Research
Council under the European Union's Seventh Framework Program (FP7/2007-2013) including ERC grant agreements 240672, 291329, and 306478.
We  acknowledge support from the Brazilian Instituto Nacional de Ci\^encia
e Tecnologia (INCT) do e-Universo (CNPq grant 465376/2014-2).

This manuscript has been authored by Fermi Research Alliance, LLC under Contract No. DE-AC02-07CH11359 with the U.S. Department of Energy, Office of Science, Office of High Energy Physics.

We are grateful for the extraordinary contributions of our CTIO colleagues and the DECam Construction, Commissioning and Science Verification
teams in achieving the excellent instrument and telescope conditions that have made this work possible.  The success of this project also 
relies critically on the expertise and dedication of the DES Data Management group.

\section*{Software}

NumPy \citep{numpy}, SciPy \citep{2020SciPy-NMeth}, pandas \citep{pandas}, Matplotlib \citep{matplotlib}, Astropy \citep{astropy:2013, astropy:2018}, h5py \citep{collette_python_hdf5_2014}. 

\section{Data Availability}
The 5 year photometric data for SN Ia and host galaxy properties from the Dark Energy Survey will be released as part of the cosmology analysis at \url{https://des.ncsa.illinois.edu/releases/sn}

\bibliographystyle{mnras}
\bibliography{bibliography} 
\newpage

\section*{Affiliations}
$^{1}$ School of Physics and Astronomy, University of Southampton,  Southampton, SO17 1BJ, UK\\
$^{2}$ Institute of Cosmology and Gravitation, University of Portsmouth, Portsmouth, PO1 3FX, UK\\
$^{3}$ Department of Astrophysics, American Museum of Natural History, Central Park West and 79th Street, New York, NY 10024–5192, USA\\
$^{4}$ Department of Physics, Carnegie Mellon University, Pittsburgh, Pennsylvania 15312, USA\\
$^{5}$ Department of Physics, Duke University Durham, NC 27708, USA\\
$^{6}$ School of Mathematics and Physics, University of Queensland,  Brisbane, QLD 4072, Australia\\
$^{7}$ Institut d'Estudis Espacials de Catalunya (IEEC), 08034 Barcelona, Spain\\
$^{8}$ Institute of Space Sciences (ICE, CSIC),  Campus UAB, Carrer de Can Magrans, s/n,  08193 Barcelona, Spain\\
$^{9}$ Centre for Gravitational Astrophysics, College of Science, The Australian National University, ACT 2601, Australia\\
$^{10}$ The Research School of Astronomy and Astrophysics, Australian National University, ACT 2601, Australia\\
$^{11}$ Fermi National Accelerator Laboratory, P. O. Box 500, Batavia, IL 60510, USA\\
$^{12}$ Department of Physics, IIT Hyderabad, Kandi, Telangana 502285, India\\
$^{13}$ Cerro Tololo Inter-American Observatory, NSF's National Optical-Infrared Astronomy Research Laboratory, Casilla 603, La Serena, Chile\\
$^{14}$ Laborat\'orio Interinstitucional de e-Astronomia - LIneA, Rua Gal. Jos\'e Cristino 77, Rio de Janeiro, RJ - 20921-400, Brazil\\
$^{15}$ Department of Physics, University of Michigan, Ann Arbor, MI 48109, USA\\
$^{16}$ CNRS, UMR 7095, Institut d'Astrophysique de Paris, F-75014, Paris, France\\
$^{17}$ Sorbonne Universit\'es, UPMC Univ Paris 06, UMR 7095, Institut d'Astrophysique de Paris, F-75014, Paris, France\\
$^{18}$ Department of Physics \& Astronomy, University College London, Gower Street, London, WC1E 6BT, UK\\
$^{19}$ Kavli Institute for Particle Astrophysics \& Cosmology, P. O. Box 2450, Stanford University, Stanford, CA 94305, USA\\
$^{20}$ SLAC National Accelerator Laboratory, Menlo Park, CA 94025, USA\\
$^{21}$ Instituto de Astrofisica de Canarias, E-38205 La Laguna, Tenerife, Spain\\
$^{22}$ Universidad de La Laguna, Dpto. Astrofísica, E-38206 La Laguna, Tenerife, Spain\\
$^{23}$ Center for Astrophysical Surveys, National Center for Supercomputing Applications, 1205 West Clark St., Urbana, IL 61801, USA\\
$^{24}$ Department of Astronomy, University of Illinois at Urbana-Champaign, 1002 W. Green Street, Urbana, IL 61801, USA\\
$^{25}$ Institut de F\'{\i}sica d'Altes Energies (IFAE), The Barcelona Institute of Science and Technology, Campus UAB, 08193 Bellaterra (Barcelona) Spain\\
$^{26}$ Jodrell Bank Center for Astrophysics, School of Physics and Astronomy, University of Manchester, Oxford Road, Manchester, M13 9PL, UK\\
$^{27}$ University of Nottingham, School of Physics and Astronomy, Nottingham NG7 2RD, UK\\
$^{28}$ Hamburger Sternwarte, Universit\"{a}t Hamburg, Gojenbergsweg 112, 21029 Hamburg, Germany\\
$^{29}$ Centro de Investigaciones Energ\'eticas, Medioambientales y Tecnol\'ogicas (CIEMAT), Madrid, Spain\\
$^{30}$ Jet Propulsion Laboratory, California Institute of Technology, 4800 Oak Grove Dr., Pasadena, CA 91109, USA\\
$^{31}$ Institute of Theoretical Astrophysics, University of Oslo. P.O. Box 1029 Blindern, NO-0315 Oslo, Norway\\
$^{32}$ Kavli Institute for Cosmological Physics, University of Chicago, Chicago, IL 60637, USA\\
$^{33}$ Department of Astronomy, University of Michigan, Ann Arbor, MI 48109, USA\\
$^{34}$ University Observatory, Faculty of Physics, Ludwig-Maximilians-Universit\"at, Scheinerstr. 1, 81679 Munich, Germany\\
$^{35}$ Santa Cruz Institute for Particle Physics, Santa Cruz, CA 95064, USA\\
$^{36}$ Center for Cosmology and Astro-Particle Physics, The Ohio State University, Columbus, OH 43210, USA\\
$^{37}$ Department of Physics, The Ohio State University, Columbus, OH 43210, USA\\
$^{38}$ Center for Astrophysics $\vert$ Harvard \& Smithsonian, 60 Garden Street, Cambridge, MA 02138, USA\\
$^{39}$ Australian Astronomical Optics, Macquarie University, North Ryde, NSW 2113, Australia\\
$^{40}$ Lowell Observatory, 1400 Mars Hill Rd, Flagstaff, AZ 86001, USA\\
$^{41}$ George P. and Cynthia Woods Mitchell Institute for Fundamental Physics and Astronomy, and Department of Physics and Astronomy, Texas A\&M University, College Station, TX 77843,  USA\\
$^{42}$ Department of Astrophysical Sciences, Princeton University, Peyton Hall, Princeton, NJ 08544, USA\\
$^{43}$ Instituci\'o Catalana de Recerca i Estudis Avan\c{c}ats, E-08010 Barcelona, Spain\\
$^{44}$ Observat\'orio Nacional, Rua Gal. Jos\'e Cristino 77, Rio de Janeiro, RJ - 20921-400, Brazil\\
$^{45}$ Department of Physics and Astronomy, Pevensey Building, University of Sussex, Brighton, BN1 9QH, UK\\
$^{46}$ Univ Lyon, Univ Claude Bernard Lyon 1, CNRS, IP2I Lyon / IN2P3, IMR 5822, F-69622 Villeurbanne, France\\
$^{47}$ Computer Science and Mathematics Division, Oak Ridge National Laboratory, Oak Ridge, TN 37831\\
$^{48}$ Lawrence Berkeley National Laboratory, 1 Cyclotron Road, Berkeley, CA 94720, USA\\
\newpage
\appendix
\section{Investigating the effects of Richness on our sample}
\label{Richness_cut_appendix}
When selecting our cluster sample we chose to use the larger $\lambda/S \geq 5$ catalogue, potentially opening our analysis to contamination from over densities of galaxies that would not be classified as true clusters. To investigate what affect this cut has, we restrict our sample to $\lambda/S \geq 15$ and re-analyse the results presented in this paper. This re-analysis is summarised in Table \ref{tab:richnesscuttable}. 

\begin{table}
\caption{Significance's of our cluster vs field analysis for a richness cut of $\lambda/S \geq 5$ and for $\lambda/S \geq 15$}
\begin{threeparttable}
\begin{tabular}{|c|c|c|}
\hline
Comparison & K-S Test result & K-S Test result \\
 & $\lambda/S \geq 5$  & $\lambda/S \geq 15$  \\
\hline
$x_{1}$  & $0.023$  & $0.0507$\\
$c$   & $0.8005$  & $0.4441$    \\
Host Stellar Mass & $0.0006$ & $0.1102$ \\
\hline
\begin{tabular}[c]{@{}l@{}}Overall Cluster Rate\\ (SNe $100\mathrm{yr}^{-1}10^{10}M_{\odot}^{-1})$\tnote{a}  \\
\end{tabular}
  &  $0.0332 ^{+0.0091}_{-0.0060} \mathrm{(syst)}$ & $0.0325 \pm 0.0098$\\
\hline
\end{tabular}
\label{tab:richnesscuttable}
\begin{tablenotes}
    \item[a] Calculated rate is for galaxies with $\mathrm{log(M_{*}/M_{\odot})}>10$ as done in Section \ref{sec:Rates}
\end{tablenotes}
\end{threeparttable}
\end{table}

Limiting the sample with a more stringent richness cut does change the significance of our results. The colour distribution changes by a large amount, but there is still no significant difference between the two samples, with both samples having less than $1\sigma$ difference between them. For the field vs cluster $x_{1}$ distribution, it does not alter the result much, with the `less significant' sample still having a tentative difference between the cluster and field samples, with a confidence level of around $95$ per cent. Additionally the restriction shifts the result into being less significant. If the cut removed non-cluster SNe this should increase the significance between our two samples, as SN $x_1$ values within rich galaxy clusters were previously found to be significantly different from field galaxy $x_1$ values, as found in \citet{Xavier}. 

\begin{figure*}
    \centering
	\includegraphics[width=\columnwidth]{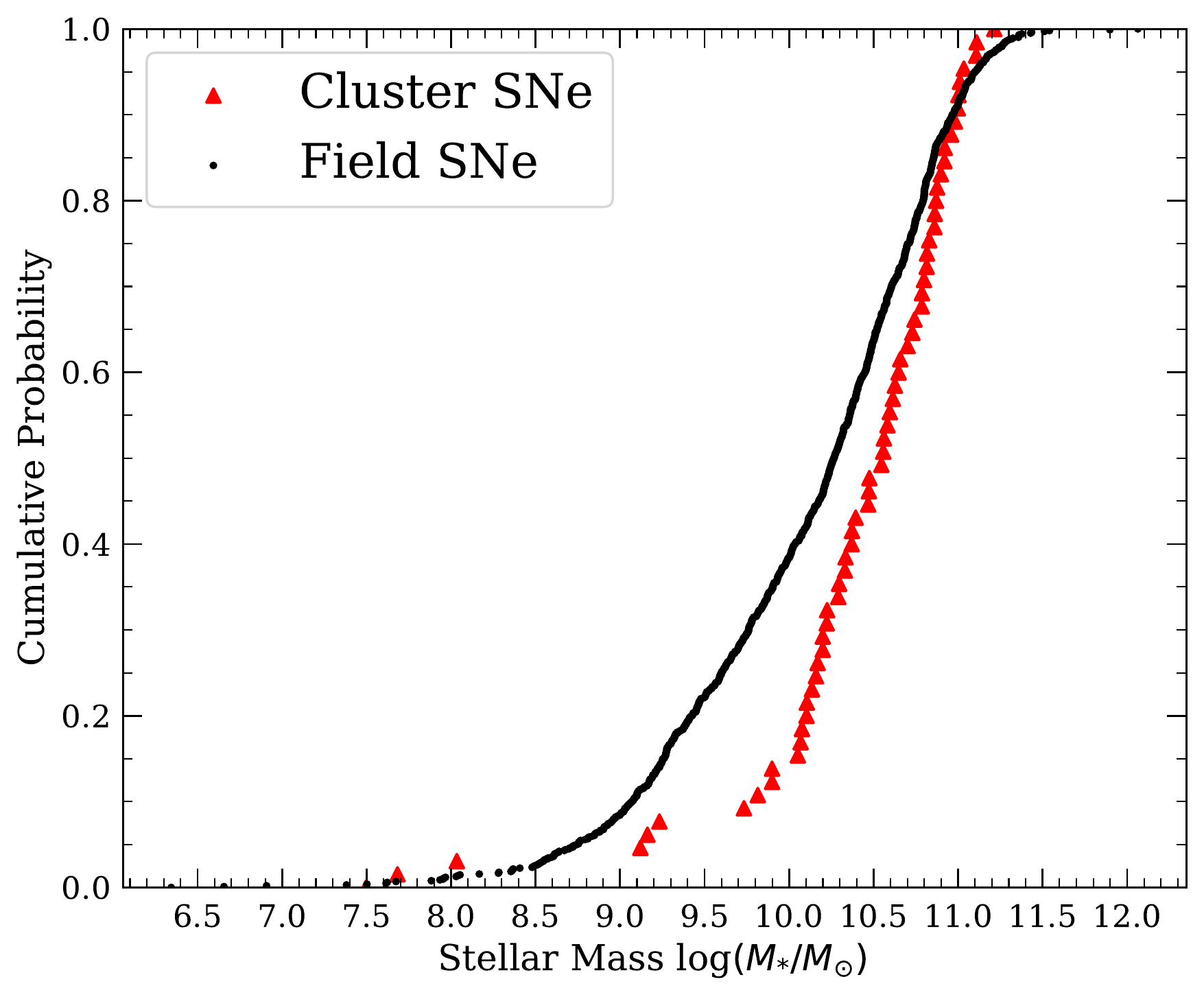}
	\includegraphics[width=\columnwidth]{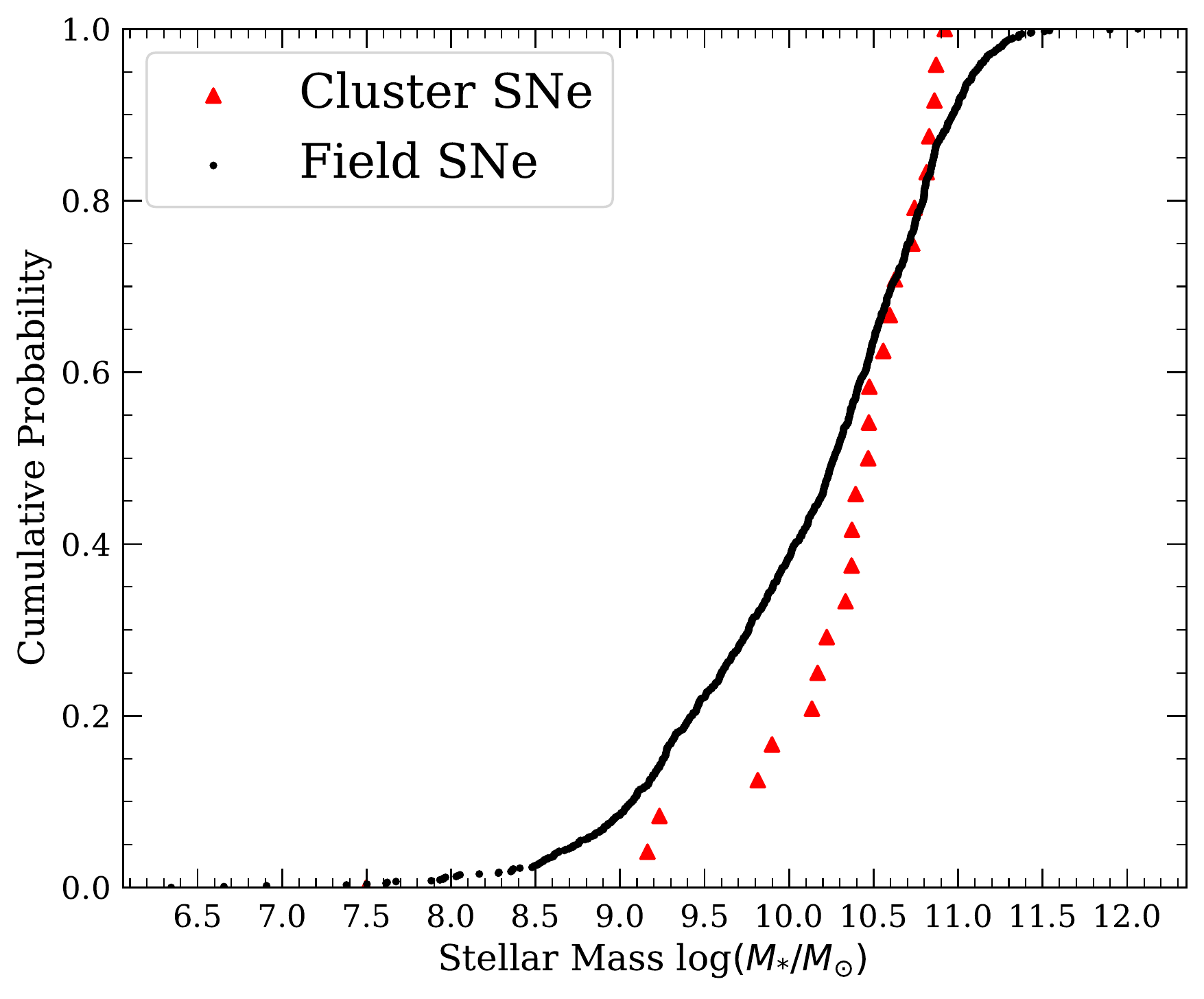}
    \caption{Host galaxy stellar mass distributions for SNe Ia occurring within clusters and in the field. Left: CDF of our uncut sample, with clusters of richness  $ \geq 5$ Right: CDF of our restricted sample, with clusters of richness  $ \geq 15$ Visually, there is little difference between the two samples, with the exception of far fewer SNe within the higher richness sample.}
    \label{fig:host_mass_appendix}
\end{figure*}

The largest shift is in the host stellar masses, where restricting our sample to only higher richness clusters again decreases the significance of the difference between them.  The actual distribution shapes however, do not significantly change. This is shown in Figure \ref{fig:host_mass_appendix}. We therefore attribute the drop in significance to lower statistics. 

There is no significant change in the overall rate, with the two being consistent when accounting for their errors. As our investigated properties do not significantly change under a richness cut, we conclude that making such a richness cut is unnecessary, and would be unnecessarily removing cluster SNe from our sample. 

\bsp
\label{lastpage}
\end{document}